**High $T_c$ superconductivity in Heavy Rare Earth hydrides: correlation between the presence of the *f* states on the Fermi surface, nesting and the value of $T_c$**

Hao Song[1], Zihan Zhang[1], Tian Cui[2,1,*], Chris J. Pickard[3,4], Vladimir Z. Kresin[5], Defang Duan[1,†]

[1]*State Key Laboratory of Superhard Materials, College of Physics, Jilin University, Changchun 130012, China*

[2]*Institute of High Pressure Physics, School of Physical Science and Technology, Ningbo University, Ningbo 315211, China*

[3]*Department of Materials Science & Metallurgy, University of Cambridge, 27 Charles Babbage Road, Cambridge CB3 0FS, United Kingdom*

[4]*Advanced Institute for Materials Research, Tohoku University 2-1-1 Katahira, Aoba, Sendai, 980-8577, Japan*

[5]*Lawrence Berkeley Laboratory, University of California at Berkeley, Berkeley, CA 94720, USA*

Corresponding authors email: [*]cuitian@nbu.edu.cn/cuitian@jlu.edu.cn
[†]duandf@jlu.edu.cn

**Abstract**

Lanthanum hydrides, containing hydrogen framework structures under compression, display a superconducting state with a high observed critical temperature. However, this phenomenon has so far only been observed at very high pressures. Here, we computationally search for superconductors with very high critical temperatures, but at much lower pressures. We uncover two such sodalite-type clathrate hydrides, $YbH_6$ and $LuH_6$ ($T_c$ =145 K at P=70 GPa for $YbH_6$ and, especially, $T_c$ =273 K at P=100 GPa for $LuH_6$). These striking properties are a consequence of the strong interrelationship between the *f* states present at the Fermi surface, structural stability and $T_c$ value. For example, $TmH_6$, with unfilled 4*f* orbitals, is stable at 50 GPa, while it has a relatively low value of $T_c$ of 25 K. As for the $YbH_6$ and $LuH_6$ compounds, they have filled *f*-shells and the decrease of the *f*-energy below the Fermi level leads to formation of the nesting regions on the Fermi surface and, as a result, to phonon "softening" and an increase in $T_c$.

It has long been predicted that monoatomic hydrogen, under high pressure, will support a transition into a metallic state, which is expected to be superconducting with a high critical temperature ($T_c$) [1]. However, for this phenomenon to be observed very high pressures are expected to be required, and it has not yet been realized [2-7]. A promising route to the realization of very high temperature conventional superconductors is the study of hydrogen-rich compounds (or hydrides); they have been shown metallize at lower pressures due to chemical precompression effect [8], see reviews [9-11]. The presence of two ions, light (H) and heavy, in the unit cell, leads to an appearance of optical high frequency phonon modes (vibrations of H ions) with a large phonon density of states (DOS) across the whole range of phonon momenta. This increases the magnitude of the electron-phonon interaction (EPI). An additional contribution due to the heavy ions (acoustic modes) further increases the EPI. Among such hydrides are $H_3S$ [12,13], $LaH_{10}$ [14,15] which high $T_c$ values were predicted and later confirmed experimentally [16-19]. These materials display the values of $T_c$ exceeding 200 K at pressures above 150 GPa. More recently, a compound of hydrogen, carbon and sulfur [20] exhibited superconductivity at 288 K, by some definitions, room-temperature, but the high pressure required (267±10 GPa) make it difficult to analyze, and rules out any practical application.

The present study is motivated by two key factors. Observed values of $T_c$ are at or close to room temperature. For example, the compound of hydrogen, carbon and sulfur, mentioned above, has $T_c \approx$ 288 K. However, thus far, the observation of high temperature superconductivity in the hydrides requires pressures as high as 267 GPa. The next frontier is the observation of room temperature superconductivity at significantly lower pressures (with a clear final goal of reaching ambient pressure). We present a step in this direction. Indeed, it will be demonstrated that high values of $T_c$ are expected to be observed in $YbH_6$ at $T_c$ =145 K and, most strikingly, in $LuH_6$ at $T_c$=273 K but at rather more moderate pressures.

Given the large number of predicted superconducting hydrides [9-11], it is now time to develop some general concepts and define major factors favorable for high values of $T_c$. Here, we focus on nesting on the Fermi surface and the corresponding role of the *f* states. Note that the sodalite cage formed by hydrogen ions was first proposed for $CaH_6$ [21]; this structural unit is present in most of the rare earth (RE) hydrides $REH_n$ (n=6, 9, 10) and appears to be beneficial for superconductivity. Yb, Lu, Tm occupy a special place among lanthanides, and we will pay particular attention to the hydrides containing these elements. Highly localized *f* electrons in RE hydrides are expected to affect superconductivity adversely, and we will discuss their impact on $T_c$ in detail below.

**Sec. I. Determination of stable stoichiometries and structural characteristics**

As is well-known, on-site Coulomb interactions are particularly significant for localized *f* electrons. As a result, the DFT+U method is usually adopted in lanthanides[22-24]. We first computed the equation of state (EOS) for $YbH_2$ and compared it with the experimental EOS to assess the reliability of our DFT calculations (see Fig.S1) [25]. The agreement with experiment is improved with the DFT+U scheme for the low-pressure phase of *Pnma*-$YbH_2$. However, ultra-soft pseudo-potentials with valence *f* electrons without a Hubbard U provided an acceptable result for the high pressure phase of $P6_3/mmc$-$YbH_2$. The authors of previous works concerning ytterbium hydrides [22,23], used a Hubbard U =3 or 5 eV for lower pressure phases and U=0 eV for high-pressure phases to reproduce available experimental data, in clear agreement with our results (see Fig.S1). Despite the absence of experiment results for $Im\bar{3}m$-$YbH_6$, our chosen pseudopotential is in excellent agreement with the full-potential treatment (see Fig. S2). Furthermore, we construct the high-pressure phase diagram of Yb-H system both without and with DFT+U (U=5 eV), through extensive structure searches via the ab initio random structure searching method as implemented in the AIRSS code [26,27]. The results are presented in Fig. S3. The convex hull is essentially unchanged for DFT+U, as compared to DFT-PBE. At high pressures, we conclude that the DFT level of description of the electronic structure is acceptable for this system.

Figure S3 shows that YbH is thermodynamically stable in the studied pressure range. So, we reconstructed the convex hull diagrams with respect to YbH and $H_2$ as shown in Fig. 1(a). Note that seven stoichiometries $YbH_2$, $Yb_2H_5$, $YbH_3$, $YbH_4$, $YbH_6$, $YbH_8$ and $YbH_{12}$ are found on the convex hull in different pressure ranges. Combining the convex hull, enthalpy differences (Fig. S4) and phonon dispersion (Fig. S5-6) of Yb-H system under high pressure, we construct the pressure-composition phase diagram in Fig. 1(b). The crystal structures of our predicted stable Yb hydrides are shown in Fig. 1(b) and Fig. S7 and the lattice parameters are provided in Table S1.

For YbH, the $Fm\bar{3}m$ phase, with hydrogens occupying octahedral interstices, transforms into a *Cmcm* phase at 170 GPa. $YbH_2$ adopts the $P6_3/mmc$ structure, which is consistent with experimental measurements [22,23]; this structure transforms into a $P6/mmm$ phase at 233 GPa. $Yb_2H_5$ has $R\bar{3}m$ symmetry between 145 and 250 GPa. $YbH_3$ favors a $Fm\bar{3}m$ structure, in which hydrogen atoms occupy both octahedral and tetrahedral interstices. $YbH_8$, with $P6_3/mmc$ symmetry, is thermodynamically favorable and dynamically stable above 240 GPa; as determined by the phonon dispersion (see Fig. S6). The high hydrogen content $YbH_{12}$, with $R\bar{3}$ symmetry, is thermodynamically stable over the pressure range of 70~240 GPa.

Structure searching for the other two heavy rare earth hydrides (HRE = Tm, Lu) adjacent to Yb, was also performed using AIRSS code [26,27]. The resulting convex hull diagrams are shown in Fig. S8. We can see that the cubic $TmH_6$ and $LuH_6$ compounds lie above the convex line, indicating that they are metastable at the corresponding pressures.

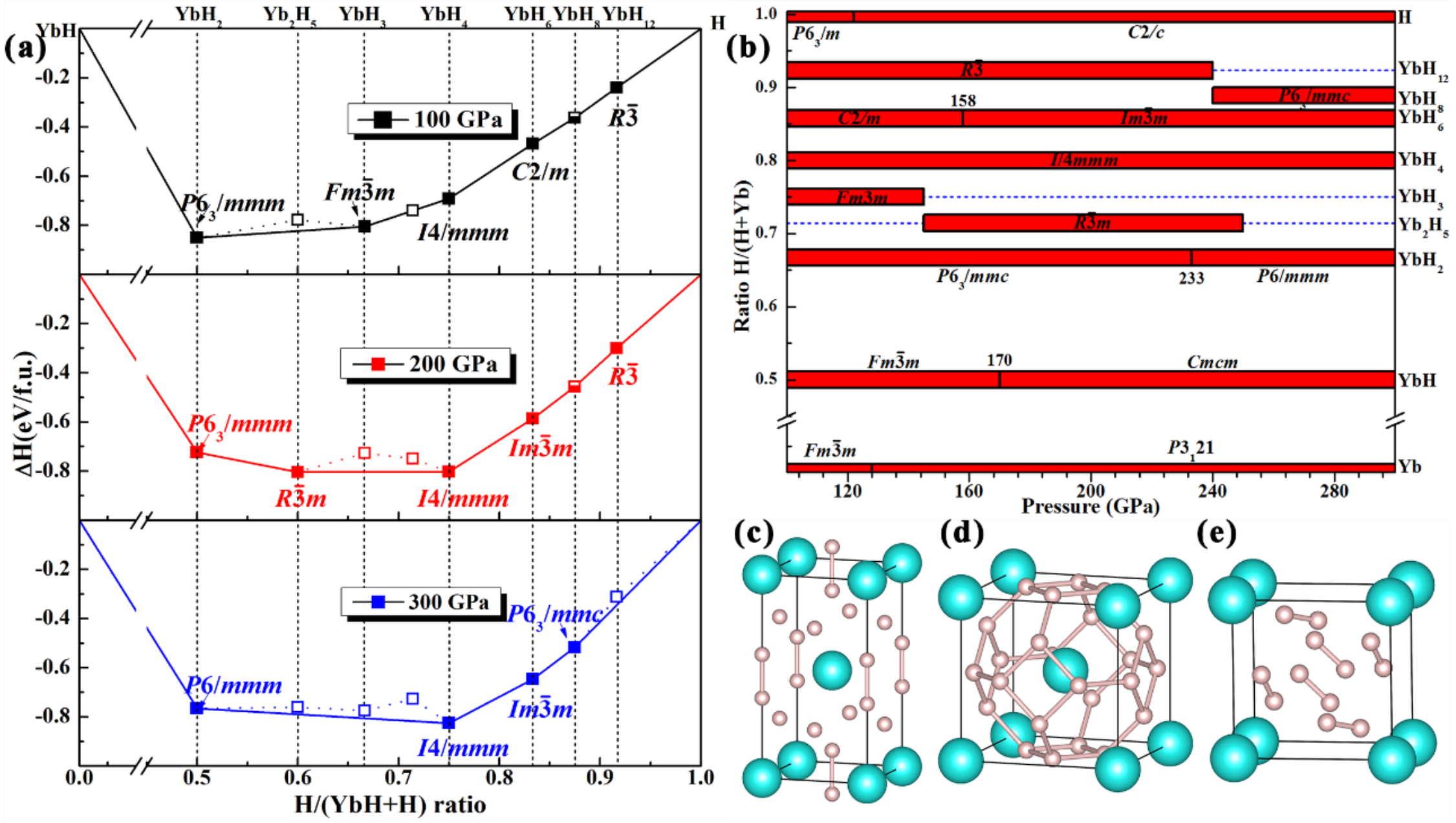

Fig.1 (a) The convex hull diagrams of Yb-H system with respect to YbH and $H_2$ at 100, 200 and 300 GPa. The solid, half filled and open square symbols indicate that the structures are thermodynamic stable, dynamic unstable, and thermodynamic unstable, respectively. (b) Pressure-composition phase diagram of Yb-H system. Crystal structures (c) $I4/mmm$-$YbH_4$, (d) $Im\bar{3}m$-$YbH_6$, and (e) $R\bar{3}$-$YbH_{12}$. The large cyan balls and small pink balls represent ytterbium and hydrogen, respectively.

**Sec. II. Analysis of electronic properties of cubic $XH_6$ (X=Tm, Yb, Lu)**

We calculated the electronic properties of $Im\bar{3}m$-$YbH_6$, as shown in Fig. 2. Remarkably, the 4*f* orbitals associated with the Yb atom form a set of localized almost non-dispersive bands that appear about 1 eV below the Fermi level. This suggests that the 4*f* orbitals will play no role in any superconductivity. There are two bands which intersect at the Fermi level, one of which is mostly occupied by *s* electrons from the H and *f* electrons from Yb, the other is occupied by *f* and *d* electrons from Yb (see Fig. S9(c, d)). The electronic density of states (DOS) at the Fermi level mainly consist of H-*s*, Yb-*f*, and Yb-*d* electrons with ratio of 25%, 50%, and 16%, respectively. Although DFT+U appears not to be necessary for the high pressure ytterbium hydrides, the value of the parameter U critically affects the electronic properties, in particular the band structure. With Hubbard U = 5 eV, we found that the *f* electrons move downward to about -2 eV and the contribution to the DOS from the *f* electrons at the Fermi level sharply decreases, while at the same time the Fermi velocity increases (see Fig.2 (b) and Fig. S9 (e, f)).

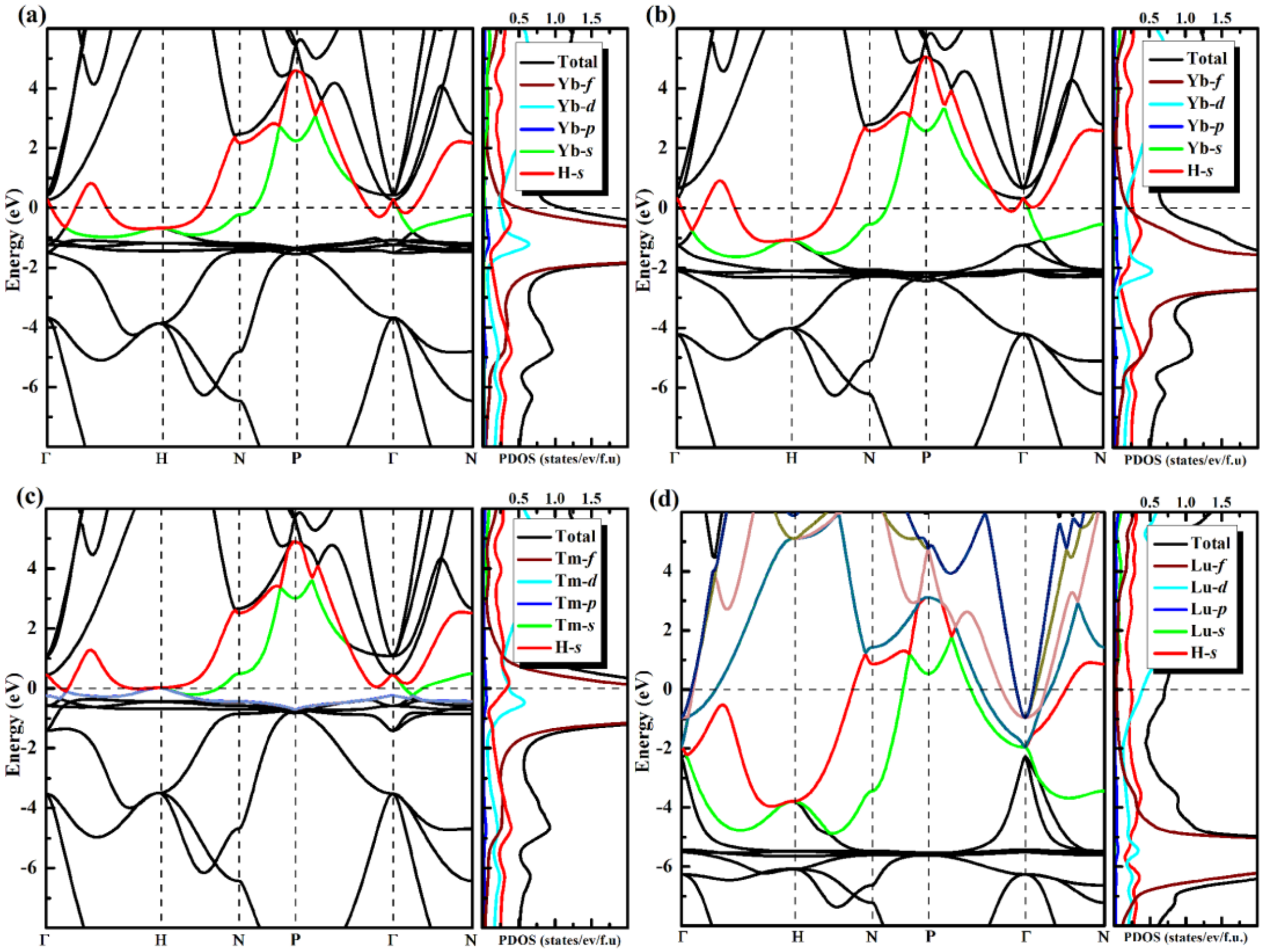

Fig.2 Band structure and projected density of electronic states (PDOS) of (a) $YbH_6$, (c) $TmH_6$ and (d) $LuH_6$ at 100 GPa. (b) Band structure and PDOS of $YbH_6$ with DFT+U (5eV).

To gain an insight into the role of the *f* electrons in the HRE sodalite-like hydrides, we calculated the band structure of $TmH_6$ and $LuH_6$. From the band structure (Fig. 3c) we can see that the *f* electrons of $TmH_6$ dominate the Fermi level with unfilled 4*f* orbitals. For $LuH_6$, the fully filled 4*f* orbitals and one extra electron in 5*d* orbitals leads to the 4*f* electrons moving downwards in the band structure (see Fig. 2d). As for $YbH_6$, electrons transfer out of the 4*f* orbitals leaving behind slightly unfilled 4*f* orbitals (0.34*e* loss), and electrons transfer from the 6*s* to partially occupy 5*d* orbitals. Therefore, the electronic configuration of $YbH_6$ is indeed inbetween that of $TmH_6$ and $LuH_6$. The location of 4*f* electron band center for $YbH_6$ does not change the number of bands crossing the Fermi level, Still, it significantly changes the Fermi velocity and the DOS of the *f* electrons at the Fermi level. Electron localization functions (ELF) of cubic $YbH_6$ suggest that a weak covalent interaction between the H atoms, with ELF of 0.58 (see Fig. S10). Yb atoms play the role of electron donors and loses approximately 1.0 *e*, and every H evenly gains 1/6 *e*.

**Sec. III. Superconducting properties of the RE hydrides.**

We now focus on the impact of the *f* states upon the superconducting properties. As was noted above, high localized *f* electrons in RE hydrides affect superconductivity adversely [15,28]. For example, experiment reported that $LaH_{10}$ and $CeH_9$ had high $T_c$ of 250-160 K and 117 K, respectively, while $T_c$ of $PrH_9$ [29] and $NdH_9$ [30] are found to be below 10 K. It is interesting that the value of the critical temperature for RE hydrides gradually decreases with increasing filling of the *f*-shell and superconductivity disappears for half-filled *f*-shells (Eu–$f^7$) [28]. One can raise an interesting question concerning the case of a filled *f*-shell. Here we report a systematic exploration of heavy rare earth hydrides (HRE= Tm, Yb, Lu) at high pressure.

Ytterbium is a unique element among the lanthanides, with a filled 4*f* orbital with configuration $4f^{14}$. It does not display ferromagnetic or antiferromagnetic order, and has a moderate crystal abundance among lanthanides. Its electronegativity is 1.1, similar to that of Ca and Y. It exhibits a mixed valence and is a superconductor at 86 GPa with $T_c$ of 1.4 K[31]. As mentioned above, we studied the various compositions of $YbH_n$. Adjacent to Yb, Tm has 13*e* filled 4*f* orbitals with $4f^{13}$, low crustal abundance among the lanthanides. Lu has a filled 4*f* orbital and an extra 5*d* electron with $4f^{14}5d^1$. It is the hardest and densest metal among lanthanides. Both dihydrdes and trihydrides of Tm and Lu exist at ambient conditions.

The most interesting hydrides at high pressure are the $TmH_6$, $YbH_6$, and $LuH_6$ compounds. They are dynamically stable at 50 GPa, 70 GPa and 100 GPa with $T_c$ of 25 K, 145 K and 273 K (ice point temperature), respectively. We uncover a non-trivial and non-monotonic dependence of $T_c$ on the degree of the *f*-shell filling. Beyond filling half of the *f*-shell, the $T_c$ of HRE hydrides gradually increases upon further filling. It reaches a maximum in the last lanthanoid hydrides $LuH_6$, which comparable with $LaH_{10}$. Importantly, the pressure of stability for $LuH_6$ is much lower than that for $LaH_{10}$ (171GPa for experimental measurements and 210 GPa for predictions at the harmonic level). This should allow the experimental observation of the superconducting state at temperatures close to room temperature, but at much lower pressure.

**Sec. IV. Spectral function and critical temperature of cubic $XH_6$ (X=Tm, Yb, Lu).**

The high $T_c$ superconducting state in the hydrides is created by strong electron-phonon coupling to high frequency optical phonons (see, e.g., [12,14,32,33]). As a first step in exploring the interactions between electron and phonon, we calculate the phonon spectrum (Fig. 3). The calculated phonon dispersion, phonon density of states for $Im\bar{3}m$ -$YbH_6$ at 70 GPa and 200 GPa, for $TmH_6$ at 50 GPa and for $LuH_6$ at 100 GPa are presented. The absence of any imaginary frequency indicates the dynamical stability of $Im\bar{3}m$-$XH_6$ (X=Tm, Yb, Lu) at their corresponding pressures. The evaluation of the phonon spectrum and the phonon DOS allows us to calculate the spectral function $\alpha^2F(\omega)$ and then the electron-phonon coupling (EPC) constant $\lambda$ determined by the relation (see Eq.S9). One can also calculate the coupling constants $\lambda_{opt}$ and $\lambda_{ac}$ describing the electron interaction with optical and acoustic phonon modes [34], see Suppl. Material [25]. As mentioned above, $Im\bar{3}m$-$YbH_6$ is dynamically stable at relatively low pressures (70 GPa). The phonon modes are separated into two parts. The acoustic phonon modes (low frequencies < 220 $cm^{-1}$) are dominated by vibrations of Yb atoms. In contrast, the vibrations of H atoms form optical phonon modes (high frequencies). For $Im\bar{3}m$-$YbH_6$, the EPC parameter $\lambda$ is 2.2 at 70 GPa, the acoustic phonon modes contribute to 10% of the total $\lambda$ (in other words, $\lambda_{ac}$ = 0.22). At 200 GPa the EPC parameter $\lambda$ is 1.00, the acoustic phonon modes contribute to 20% of the total $\lambda$ (in other words, $\lambda_{ac}$ = 0.2). Note that the strength of the electron-phonon coupling from the acoustic phonon modes of Yb atom is almost a constant with variation of pressure.

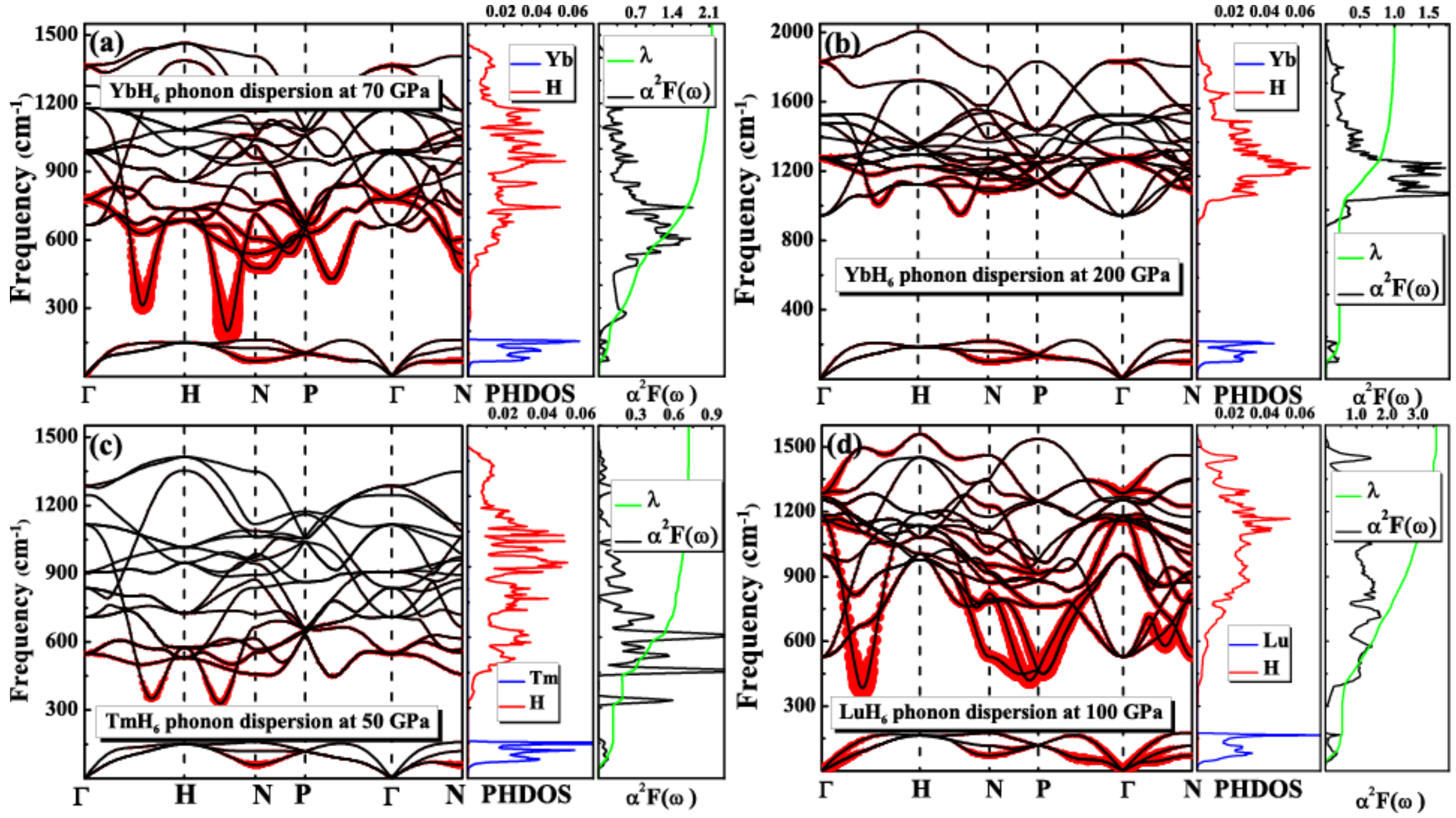

Fig.3 Phonon dispersion, phonon density of state, spectral function ($\alpha^2F(\omega)$) and integral EPC $\lambda$ of $Im\bar{3}m$-$YbH_6$ (a) at 70 and (b) 200 GPa, $TmH_6$ (c) at 50 GPa, $LuH_6$ (d) at 100 GPa. Red circles with radius proportional to the EPC strength.

A striking feature of the phonon dispersion of the cubic $YbH_6$ is the presence of soft phonon modes along Γ-H and H-N directions, as shown in Fig. 3 (a). With increasing pressure from 70 to 100 GPa, the scale of such phonon "softening" rapidly decreases and the EPC parameter $\lambda$ drastically decreases from 2.2 to 1.3 , which leads to a rapid reduction of $T_c$ from 145 to 116 K (see Fig S11). To explain the phonon "softening", we performed the calculations of Fermi surface nesting function. It is shown that there are strong nesting along Γ-H and H-N directions corresponding to where the phonon "softening" appear (see Fig. S12). The maximum $T_c$ of 145 K at 70 GPa is related to the presence of soft modes in the phonon dispersion induced by Fermi surface nesting that lead to strong electron-phonon coupling. Furthermore, combining with three-dimensional Fermi surface (Fig. S9), we found parts of the Fermi surface with strong nesting contributed by hydrogen *s* states.

To gain more insight into correlation between the presence of the *f* states on the Fermi surface, stability and the value of $T_c$, we simulate the phonon dispersion and electron-phonon coupling with an extreme case of the *f* manifold frozen in the Yb pseudopotential (see Fig. S13). The frozen-*f* calculations show that the cubic structure becomes dynamic stable just only above 100 GPa and the EPC parameter $\lambda$ is about 5.5 at 100 GPa and $T_c$=262 K with $\mu^*$=0.1. An increase in pressure leads to sharp decrease in the value of the EPC parameter λ (λ= 2.2 at 200 GPa), but λ is still more than twice as large as that in the presence of *f* electrons. Therefore, without *f* electrons at Fermi level, the scale of phonon "softening" increase, the strength of the electron-phonon coupling and $T_c$ increase, but the dynamical stablity reduces.

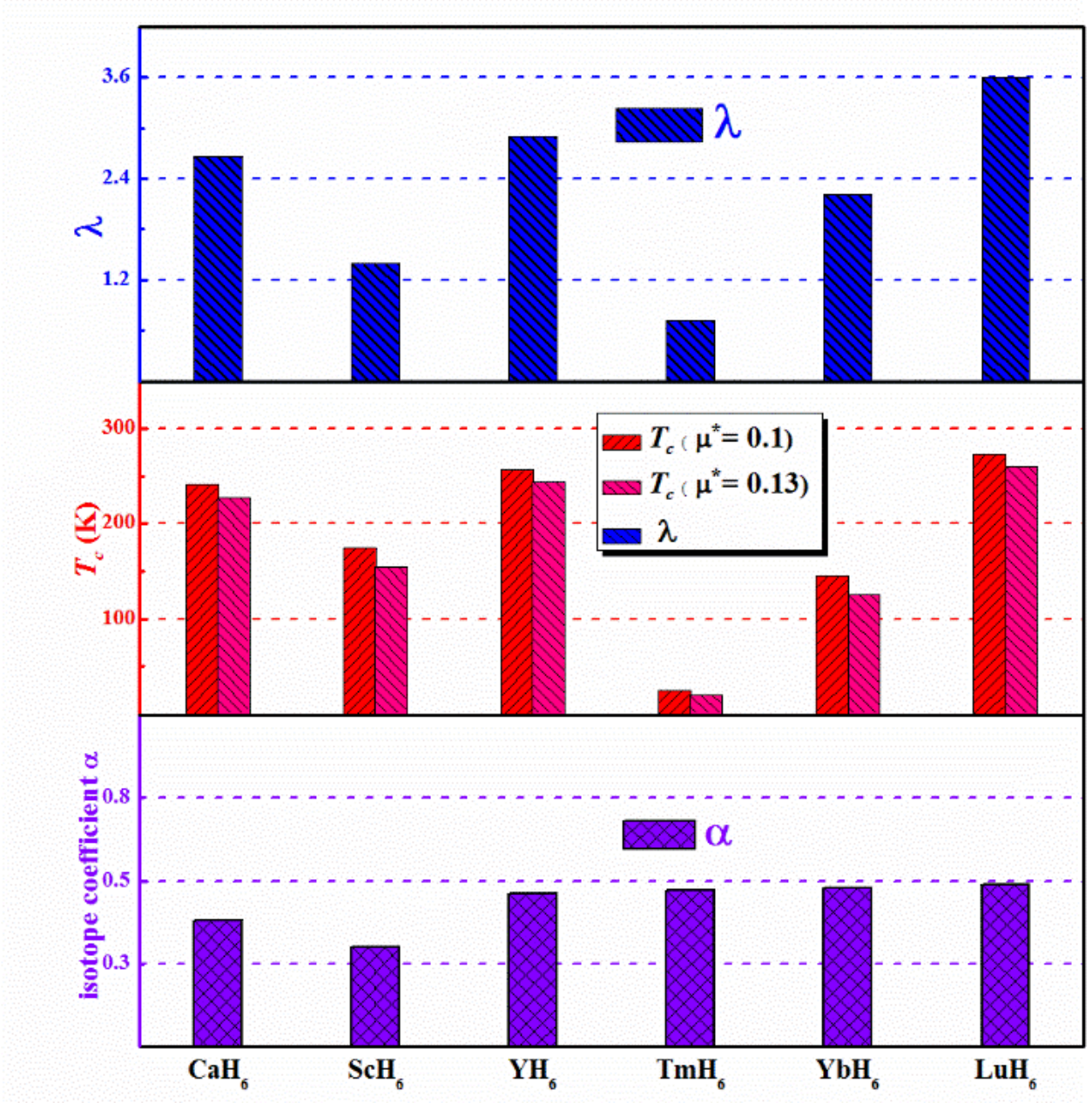

Fig.4 Superconductivity critical temperature $T_c$ and EPC parameter λ of various clathrate hydrides at high pressures. The pressure values are 150, 300, 120, 100, 50, 70 and 100 GPa for $CaH_6$, $ScH_6$, $YH_6$, $LaH_6$, $TmH_6$ $YbH_6$, $LuH_6$, respectively.

From the results described above, one can conclude that the presence of the *f* electrons in valence state leads to suppression of the phonon "softening". The stability of cubic $YbH_6$ is enhanced, but the strength of the electron-phonon coupling is reduced. Therefore, the presence of the *f* electronic states on the Fermi surface increases the dynamical stability but negatively impacts superconductivity. To further demonstrate this, phonon dispersion and EPC parameter λ of $TmH_6$ and $LuH_6$ are also calculated (see Fig.3(c, d)). It is shown that $TmH_6$ is stable at 50 GPa with λ = 0.72 and $T_c$ = 25 K. In sharp contrast, $LuH_6$ is stable at 100 GPa with λ = 3.60 and $T_c$ =273 K. In each case, the EPC parameter λ decreases with increasing pressure. The values of $T_c$ for $TmH_6$, $YbH_6$ and $LuH_6$, mentioned above, are estimated self-consistent solving the Eliashberg equations (scE), as shown in Fig. S14. We also calculated the values of the critical temperature using the Allen−Dynes-modified McMillan equation (A-D-M) and Gor'kov-Kresin equation (G-K) equation; the obtained values of $T_c$ appear to be slightly lower than those obtained with scE, see Table. S4.

Except for lanthanum and cerium, the light lanthanoids do not look being promising for high $T_c$ superconductivity, and this is due to the presence of the extra *f* electrons, which affect superconductivity adversely. But as for the heavy lanthanide, ytterbium, with its filled 4*f* orbitals, the critical temperature of $YbH_6$ compound is high relative to $CeH_9$ ($T_c$= 117 K, Ce has 1*e* filled 4*f* orbitals).

The hydride $LuH_6$ with filled 4*f* orbitals and 1*e* filled 5*d* valence orbitals appears to be a remarkable compound with the highest $T_c$ among other cubic $XH_6$; its $T_c$ value is comparable with that for the $LaH_{10}$ ($T_c$= 250-260 K, La has 1*e* filled 5*d* orbitals). Moving from the light to heavy lanthanoids, one can observe that the $T_c$ value initially decreases then gradually increases in a "secondary wave" upon further filling of the *f*-shells above its half-filling. Finally, it reaches a maximum $T_c$ for the last lanthanoid hydride, $LuH_6$. It has been a long term goal to obtain high-temperature superconductors in hydrides at pressures below 100 GPa. Here, we report the realization of this dream, for $YbH_6$ and $LuH_6$ compounds.

As mentioned above, several megabar pressures currently required to force the hydrides into the superconducting states to get high $T_c$. But there should be a compromise and balance between the pressure required and $T_c$. There has been proposed a figure of merit *S* (see Eq.8 in [35]), which makes the compromise explicit. Lower values of both $S(H_3S)$ and $S(LaH_{10})$ equal to 1.3 reflect the very high pressures required to achieve the superconducting states. As for our results, $S(YbH_6)$ = 1.8 and $S(LuH_6)$ = 2.5, both exceed the values of $S(H_3S)$ and $S(LaH_{10})$. When we plot these *S* values and pressures in figure of merit *S* (see Fig. 10 in [34]), we find that $LuH_6$ sits somewhere between the iron based superconductors and captures.

The value of the isotope coefficient for the H→D substitutions relates directly to the interplay of the optical and acoustic phonon modes. We calculated the isotope coefficient for cubic $XH_6$ hydrides (X=Yb, Lu, Tm, Ca, Sc, Y). The two-coupling constants method [34] was employed (see Eq.S10). One can see that the isotope coefficients for cubic lanthanide hydrides are large and close to the optimum value ($\alpha_{max}$=0.5). Such large values reflect the dominant contribution of the optical modes (see Table S4). As for $CaH_6$ and $ScH_6$, the values of isotope coefficients are also large, but smaller than for $YbH_6$ and $LuH_6$; they are comparable with those for $H_3S$, see [34,36]. The critical temperature of $XD_6$ ($T_c^D$) was estimated according to the equation of $T_c/T_c^D = (M_D/M_H)^\alpha$, where $T_c$ is obtained from the G-K equation (see Table S5). The expression for the isotope coefficient (see Eq.S20) was derived from the G-K equation (see Eq.S10) and is based on the two coupling constants method.

In conclusion, we investigated the heavy rare earth hydrides (HRE= Tm, Yb, Lu), especially those with the sodalite hydrogen cage structure, at high pressure. The temperature of the transition into the superconducting state correlates strongly with the presence of *f* states at the Fermi level. This is manifested in the nesting regions existence on the Fermi surface and corresponds to the appearance of phonon "softening". For example, the $TmH_6$ compound is stable at low pressure (50 GPa), but the $T_c$ is relatively low ($T_c$=25 K). One can observe much higher $T_c$ for $YbH_6$ ($T_c$= 145 K at 70 GPa). The optimal case corresponds to $LuH_6$ hydride. Remarkably the critical temperature for this compound can reach $T_c$=273 K at P=100 GPa. Such a pressure is much below those required for other high $T_c$ hydrides. One can speculate that cubic clathrate ternary hydrides doped with heavy rare earth elements Yb and Lu have the potential to be synthesized at much lower pressure and at same time exhibit high temperature superconductivity.

**Acknowledgement**

This work was supported by National Natural Science Foundation of China (Nos. 51632002, 11674122 and 11974133), Program for Changjiang Scholars and Innovative Research Team in University (No. IRT_15R23). C.J.P. acknowledges financial support from the Engineering and Physical Sciences Research Council (Grant EP/P022596/1). Parts of the calculations were performed in the High Performance Computing Center of Jilin University and TianHe-1(A) at the National Supercomputer Center in Tianjin.

# Supplementary information

## High $T_c$ superconductivity in heavy Rare Earth Hydrides: correlation between the presence of the f states on the Fermi surface, nesting and the value of $T_c$

Hao Song[1], Zihan Zhang[1], Tian Cui[2,1,*], Chris J. Pickard[3,4,*], Vladimir Z. Kresin[5], Defang Duan[1,†]

[1]*State Key Laboratory of Superhard Materials, College of Physics, Jilin University, Changchun 130012, China*

[2]*Institute of High Pressure Physics, School of Physical Science and Technology, Ningbo University, Ningbo 315211, China*

[3]*Department of Materials Science & Metallurgy, University of Cambridge, 27 Charles Babbage Road, Cambridge CB3 0FS, United Kingdom*

[4]*Advanced Institute for Materials Research, Tohoku University 2-1-1 Katahira, Aoba, Sendai, 980-8577, Japan*

[5]*Lawrence Berkeley Laboratory, University of California at Berkeley, Berkeley, CA 94720, USA*

Corresponding authors email: *cuitian@nbu.edu.cn/cuitian@jlu.edu.cn

†duandf@jlu.edu.cn

# Content

## Computational method

For the structure searching we used the ab initio random structure searching (AIRSS) approach, [1,2], combination with the density functional theory electronic structure code CASTEP [3-5]. The exchange-correlation functional was described using Perdew–Burke–Ernzerhof (PBE) of generalized gradient approximation (GGA) [6]. The plane-wave cutoff of 400 eV and Monkhorst-Pack[7] k-point spacing of $2\pi\times0.07$ Å$^{-1}$ were chose. The ultrasoft pseudopotentials [8] with the valence electrons Yb ($5s^2 5p^6 4f^{14} 6s^2$), Tm ($5s^2 5p^6 4f^{13} 6s^2$), Lu ($5s^2 5p^6 4f^{14} 5d^1 6s^2$) and $1s^1$ for H were used. Extensive searches were performed at 100, 200 and 300 GPa with 1 to 24 hydrogen atoms and 1- 4 ytterbium atoms, which yielded about 7000 structures per search. . . Then, we re-optimized the structures on the convex hull using CASTEP code with high quality parameters to re-confirm the convex hull. The same structure searching for Tm-H and Lu-H system were also performed by AIRSS code as Yb-H system. The on-site Coulomb repulsion among the localized Yb 4*f* electrons was described by the DFT+U method, with U = 5 eV. Phonon and electron–phonon coupling (EPC) were calculated in the framework of density-functional perturbation theory implemented in the Quantum ESPRESSO package[9]. Ultrasoft pseudopotentials were used with a kinetic energy cutoff of 80 Ry. We adopted $18\times18\times18$ k-points grid and $6\times6\times6$ q-points for $Im\bar{3}m$-$YbH_6$. The superconductive transition temperatures $T_c$ are evaluated with use of the Eliashberg equation [10] with Coulomb pseudopotential $\mu^*$=0.1-0.13. We employed three different methods: the Allen−Dynes-modified McMillan equation (A-D-M) with correction factors [11]; Self-consistent solution of the Eliashberg equations (scE) [11-13]; Gor’kov-Kresin equation (G-K) [14].

## EOS of $YbH_2$ and $YbH_6$

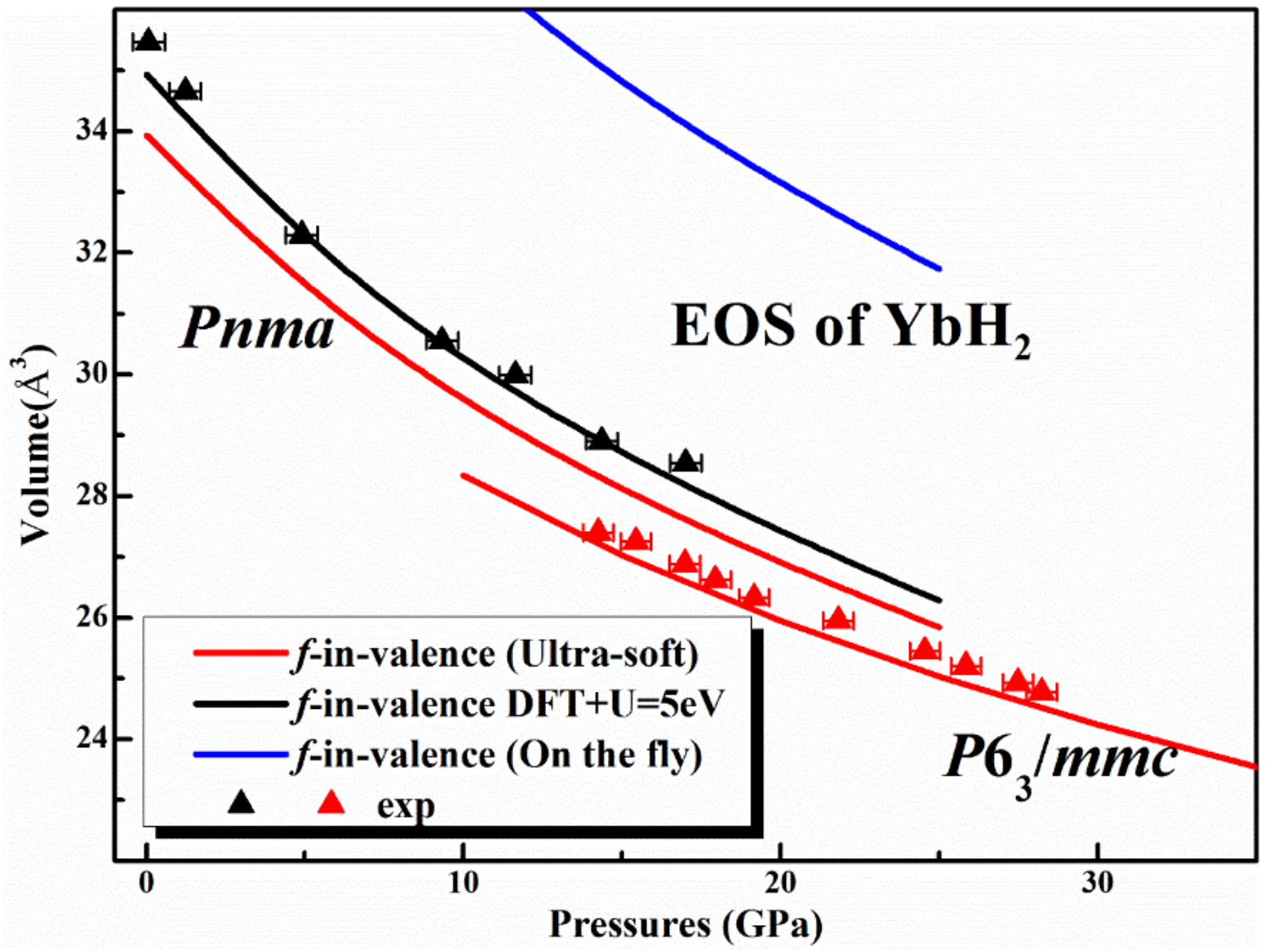

**Fig. S1.** Volumes as a function of pressures for $YbH_2$ calculated by using ultra-soft potentials in CASTEP. The agreement with experiment is improved for *Pnma* phase of $YbH_2$ with DFT+U=5 eV. High pressure phase of $P6_3/mmc$-$YbH_2$ gave identical results without DFT+U.

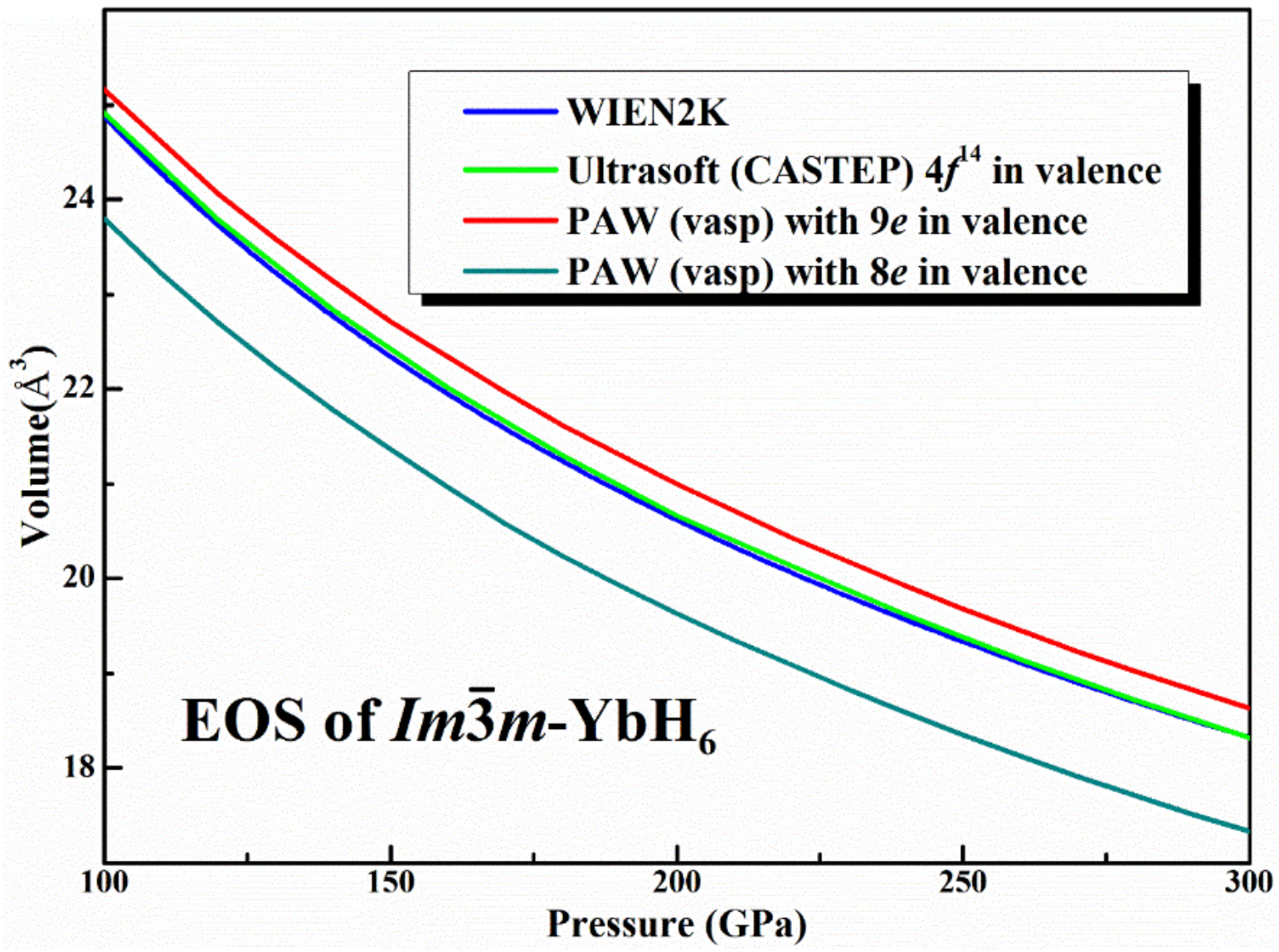

**Fig. S2.** Volumes as a function of pressures for $YbH_6$ calculated by using full-potential in WIEN2K, ultra-soft potentials in CASTEP and PAW potential in VASP. Compared with full-potential results, ltra-soft potentials in CASTEP gave identical results, but VASP calculations deviate.

## Convex hull diagram, structural parameters, enthalpy difference curves and phonon dispersion of Yb-H

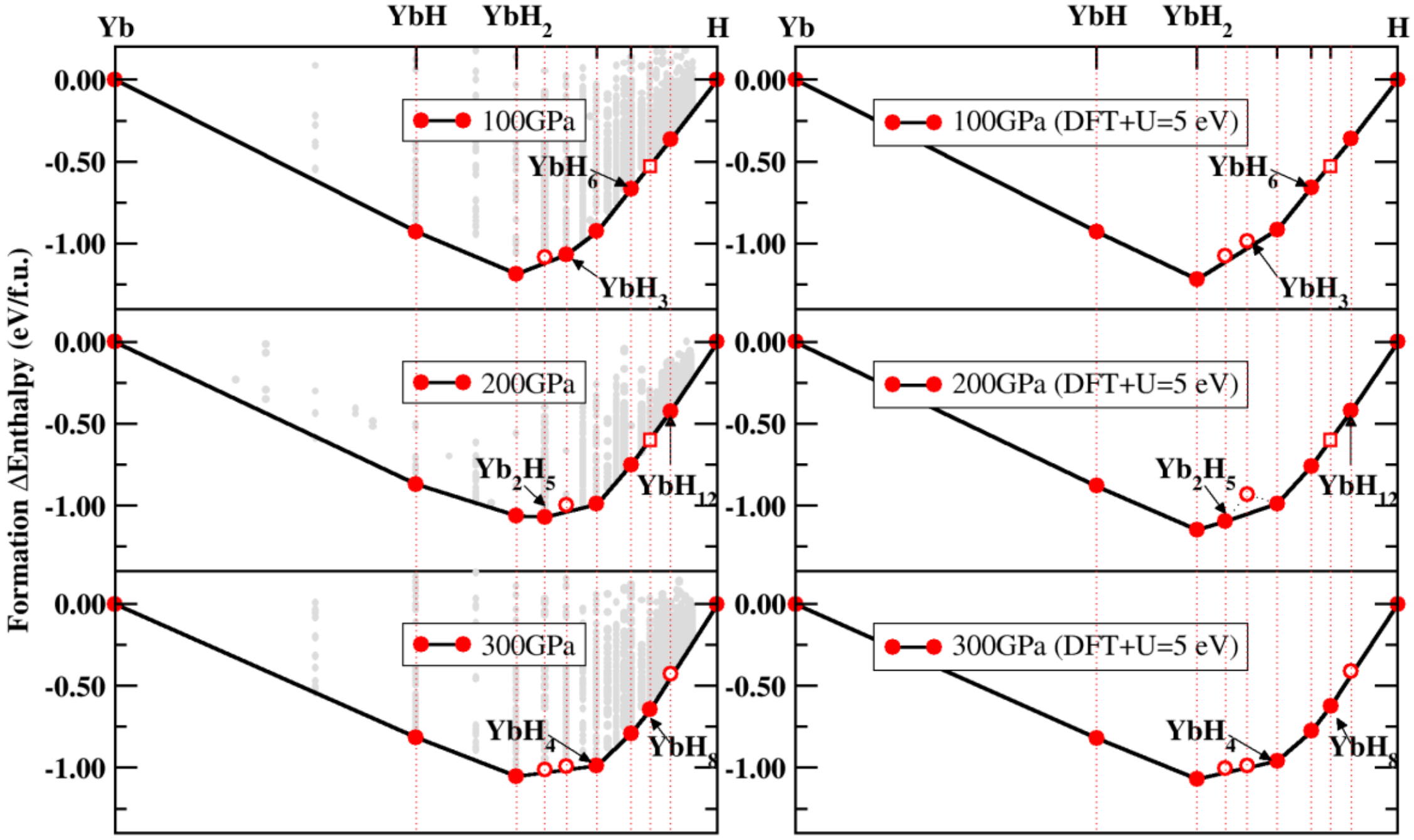

**Fig. S3.** Phase diagrams of the Yb-H without and with DFT+U (U=5 eV) at 100, 200 and 300 GPa, respectively. The solid circle, open circle and open square symbols indicate that the structures are thermodynamic stable, thermodynamic unstable and dynamic unstable, respectively. As can be seen, DFT+U slightly changes the convex-hull. With DFT+U, $YbH_3$ is off the convex hull at 100 GPa, without DFT+U $YbH_3$ is off the convex hull at 200 GPa.

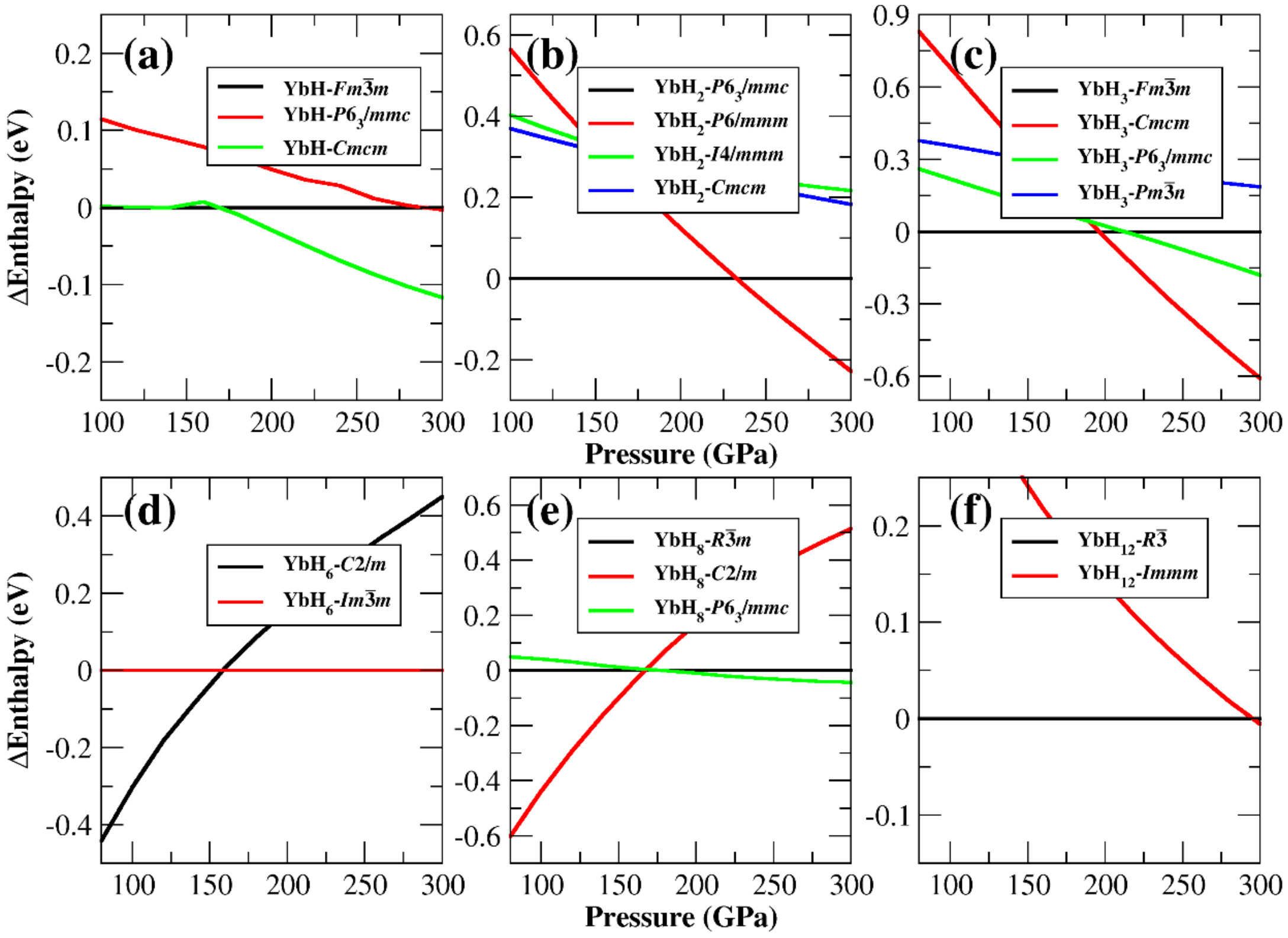

**Fig. S4.** The enthalpy difference ΔH of $YbH_n$ (n=1, 2, 3, 6, 8 and 12) from 100 to 300 GPa.

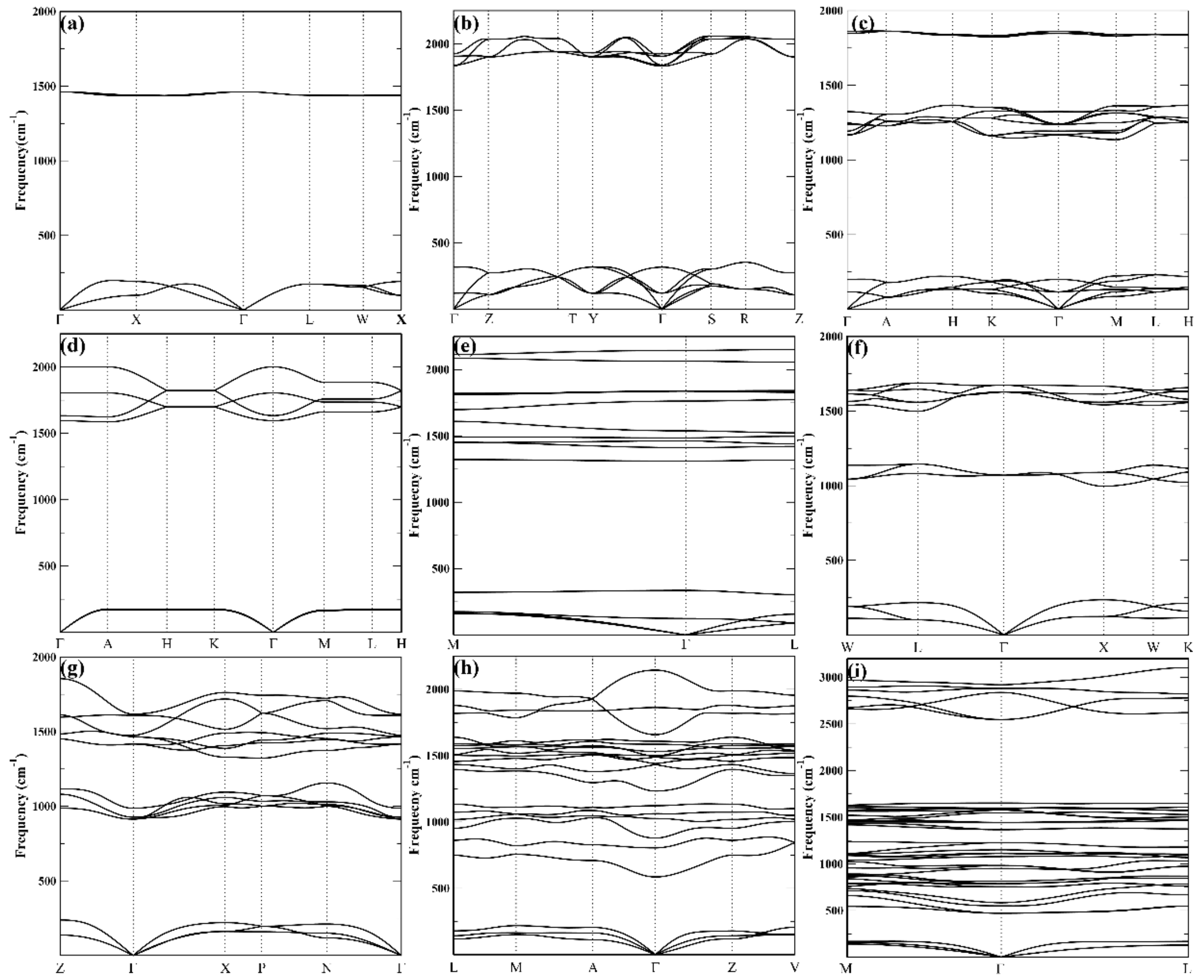

**Fig. S5.** Phonon dispersion of (a) *Fm*$\bar{3}$*m*-YbH at 100 GPa, (b) *Cmcm*-YbH at 250 GPa, (c) *P*6$_3$/*mmc*-$YbH_2$ at 100 GPa, (d) *P*6/*mmm*-$YbH_2$ at 300 GPa, (e) *R*$\bar{3}$*m*-$Yb_2H_5$ at 200 GPa, (f) *Fm*$\bar{3}$*m*-$YbH_3$ at 100 GPa, (g) *I*4/*mmm*-$YbH_4$ at 100 GPa, (h) *C*2/*m* $YbH_6$ at 100 GPa, (i) *R*$\bar{3}$-$YbH_{12}$ at 100 GPa.

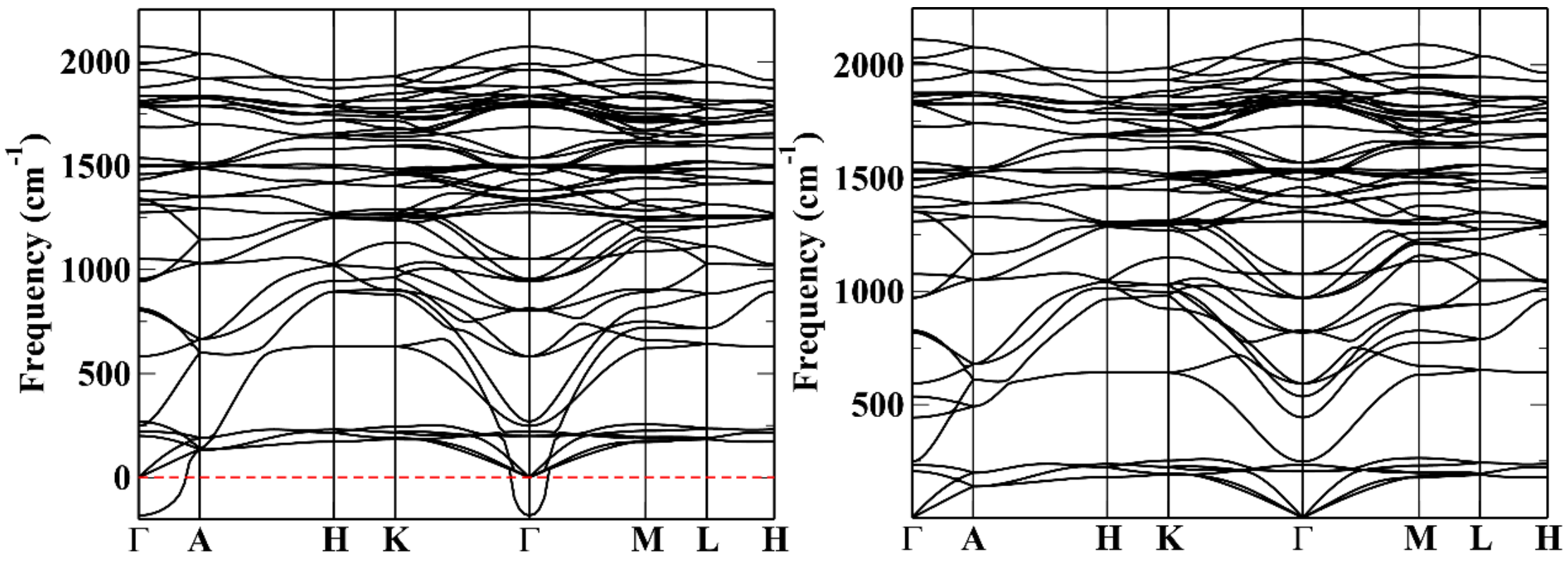

**Fig. S6.** Phonon dispersion of *P*6$_3$/*mmc*-$YbH_8$ at 220 GPa (left) and 240 GPa (right).

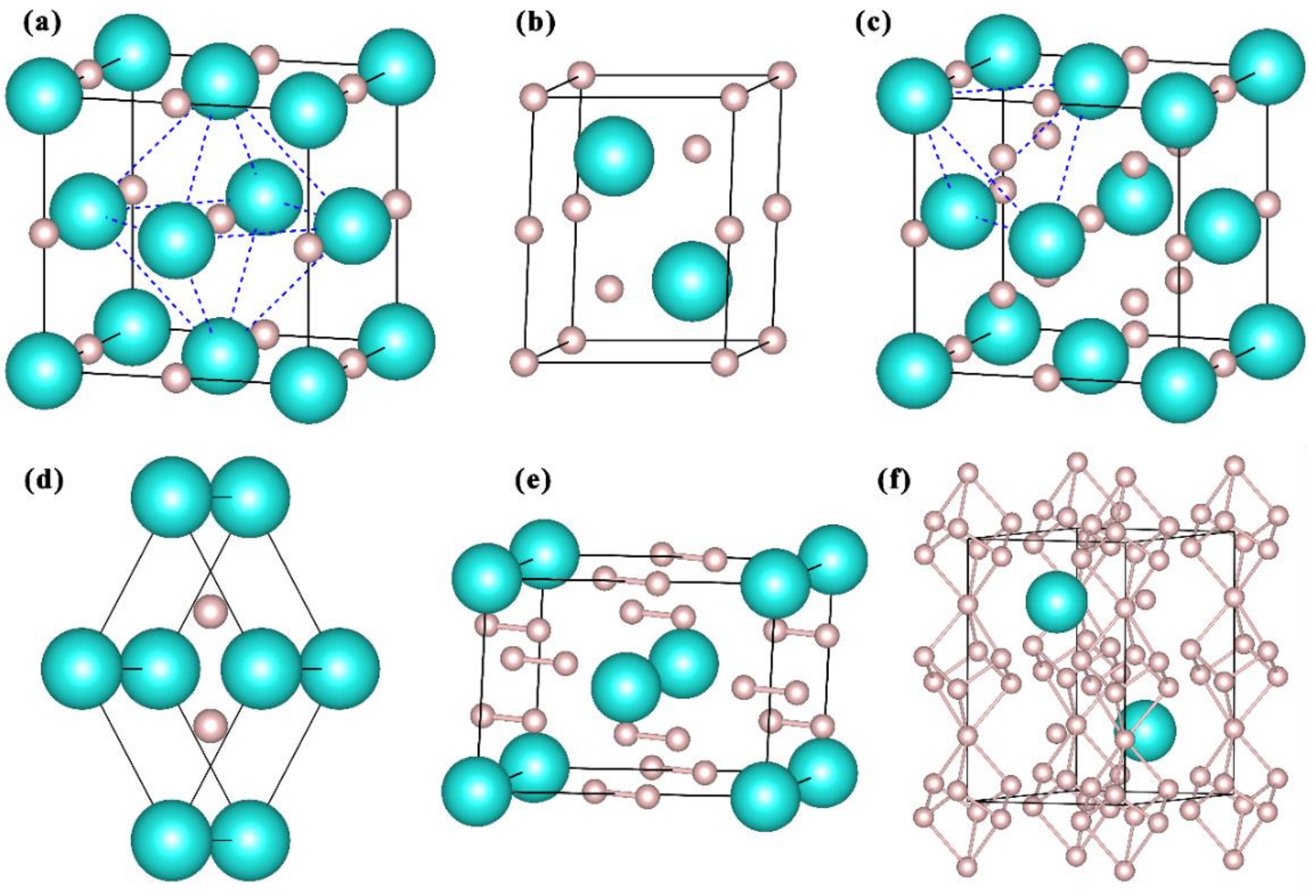

**Fig. S7.** The structure of Yb hydrides (a) $Fm\bar{3}m$-YbH (b) $P6_3/mmc$-$YbH_2$ (c) $Fm\bar{3}m$-$YbH_3$ (d) $P6/mmm$-$YbH_2$ (e) $C2/m$-$YbH_6$ (with H-H distance about 1.0 Å) (f) $P6_3/mmc$-$YbH_8$ (with H-H distance about 1.32 Å).

**Table S1.** Calculated structural parameters of stable $YbH_n$ (n=1-12) compounds.

| | Space group | Lattice parameters (Å) | Atoms | Atomic coordinates (fractional) X | Y | Z |
|---|---|---|---|---|---|---|
| YbH<br>(100 GPa) | $Fm\bar{3}m$ | $a=b=c=$ 4.01<br>$\alpha=\beta=\gamma=90°$ | H(4b)<br>Yb(4a) | 0.50<br>0.00 | 0.50<br>0.00 | 0.50<br>0.00 |
| YbH<br>(200 GPa) | $Cmcm$ | $a=$ 3.74<br>$b=$ 3.88<br>$c=$ 3.57<br>$\alpha=\beta=\gamma=90°$ | H(4c)<br>Yb(4c) | 0.50<br>0.50 | 0.33<br>0.20 | 0.25<br>0.75 |
| $YbH_2$<br>(100 GPa) | $P6_3/mmc$ | $a=b=$3.22<br>$c=$ 4.10<br>$\alpha=\beta=\gamma=90°$ | H(2d)<br>H(2a)<br>Yb(2c) | 1/3<br>0.00<br>2/3 | 2/3<br>0.00<br>1/3 | 0.75<br>0.00<br>0.75 |
| $YbH_2$<br>(300 GPa) | $P6/mmm$ | $a=b=$2.52<br>$c=$ 2.33<br>$\alpha=\beta=90°$<br>$\gamma=120°$ | H(2d)<br>Yb(1a) | 2/3<br>0.00 | 1/3<br>0.00 | 0.50<br>0.00 |
| $YbH_3$<br>(100 GPa) | $Fm\bar{3}m$ | $a=b=c=$ 4.36<br>$\alpha=\beta=\gamma=90°$ | H(8c)<br>H(4b)<br>Yb(4a) | 0.25<br>0.50<br>0.00 | 0.25<br>0.50<br>0.00 | 0.25<br>0.50<br>0.00 |
| $YbH_4$ (120 GPa) | *I4/mmm* | $a$=$b$=2.84<br>$c$ = 5.12<br>$\alpha$ =$\beta$ =$\gamma$ =90° | H(4d)<br>H(4e)<br>Yb(2a) | 0.50<br>0.50<br>0.00 | 0.00<br>0.50<br>0.00 | 0.25<br>0.094<br>0.00 |
| $YbH_6$ (100 GPa) | $C2/m$ | $a$ = 5.14<br>$b$ = 3.80<br>$c$ = 3.04<br>$\alpha$ =$\gamma$ =90° $\beta$ =119° | H(8j)<br>H(4i)<br>Yb(2a) | 0.11<br>0.61<br>0.00 | -0.29<br>0.00<br>0.00 | 0.59<br>0.09<br>0.00 |
| $YbH_6$ (70 GPa) | $Im\bar{3}m$ | $a$=$b$=$c$=3.79<br>$\alpha$ =$\beta$ =$\gamma$ =90° | H(12d)<br>Yb(2a) | 0.25<br>0.00 | 0.00<br>0.00 | 0.50<br>0.00 |
| $YbH_6$ (200 GPa) | $Im\bar{3}m$ | $a$=$b$=$c$=3.47<br>$\alpha$ =$\beta$ =$\gamma$ =90° | H(12d)<br>Yb(2a) | 0.25<br>0.00 | 0.00<br>0.00 | 0.50<br>0.00 |
| $YH_6$ [15] (200 GPa) | $Im\bar{3}m$ | $a$=$b$=$c$=3.53<br>$\alpha$ =$\beta$ =$\gamma$ =90° | H(12d)<br>Yb(2a) | 0.25<br>0.00 | 0.00<br>0.00 | 0.50<br>0.00 |
| $CaH_6$ [16] (150 GPa) | $Im\bar{3}m$ | $a$=$b$=$c$=3.50<br>$\alpha$ =$\beta$ =$\gamma$ =90° | H(12d)<br>Yb(2a) | 0.25<br>0.00 | 0.00<br>0.00 | 0.50<br>0.00 |
| $YbH_8$ (300 GPa) | $P6_3/mmc$ | $a$=$b$=3.10<br>$c$= 5.07<br>$\alpha$ =$\beta$ =90°<br>$\gamma$ =120° | H(12k)<br>H(2d)<br>H(2b)<br>Yb(2c) | 0.16<br>2/3<br>0.00<br>2/3 | 0.33<br>1/3<br>0.00<br>1/3 | 0.94<br>0.25<br>0.25<br>0.75 |
| $YbH_{12}$ (200 GPa) | $R\bar{3}$ | $a$=$b$=4.44<br>$c$=5.52<br>$\alpha$ =$\beta$ = 90°<br>$\gamma$ =120° | H(18f)<br>H(18f)<br>Yb(3a) | 0.24<br>-0.17<br>0.00 | 0.25<br>-0.71<br>0.00 | -0.56<br>0.03<br>0.00 |

**Table S2.** Distance of molecular $H_2$ in $XH_4$ (X=Yb, Y, Ca) and $C2/m$-$YbH_6$. Distance of H-H in $Im\bar{3}m$-$YbH_6$.

| Hydrides / Pressure =120 GPa | $CaH_4$ | $YbH_4$ | $YH_4$ [17] | $C2/m$-$YbH_6$ | $Im\bar{3}m$-$YbH_6$ |
|---|---|---|---|---|---|
| | 0.82 Å | 0.96 Å | 1.33 Å | 0.96 ~ 0.99 Å | 1.28 Å |

$YbH_4$ maintains the space group *I4/mmm* isotypic with $SrH_4$ [18], $CaH_4$ [16] and $YH_4$ [17] which contains monatomic H and molecular $H_2$. The H-H distance of $H_2$ unit in $YbH_4$ is longer than $CaH_4$, shorter than $YH_4$, as shown in Table S2. $YbH_6$ has two phases of $C2/m$ and $Im\bar{3}m$, and the phase transition occurs at 158 GPa. The $C2/m$ phase of $YbH_6$ also contain molecular $H_2$ and its H-H distance is comparable with $H_2$ unit in $YbH_4$ at same pressure (see Table S2). The $Im\bar{3}m$ phase contains clathrate $H_{24}$ cage consist of eight $H_6$ hexagons and six $H_4$ squares. The H-H distance of $Im\bar{3}m$ phase is 1.28 Å at 120 GPa which is substantially longer than the $H_2$ unit distance in $C2/m$ phase. The H-H distance of $H_2$ correlate negatively with bonding strength. From $H_2$ bonding strength, we can conjecture that Yb contributed more electrons to $H_2$ unit than Ca but less than Y at high pressure.

## Convex hull diagram of Lu-H and Tm-H

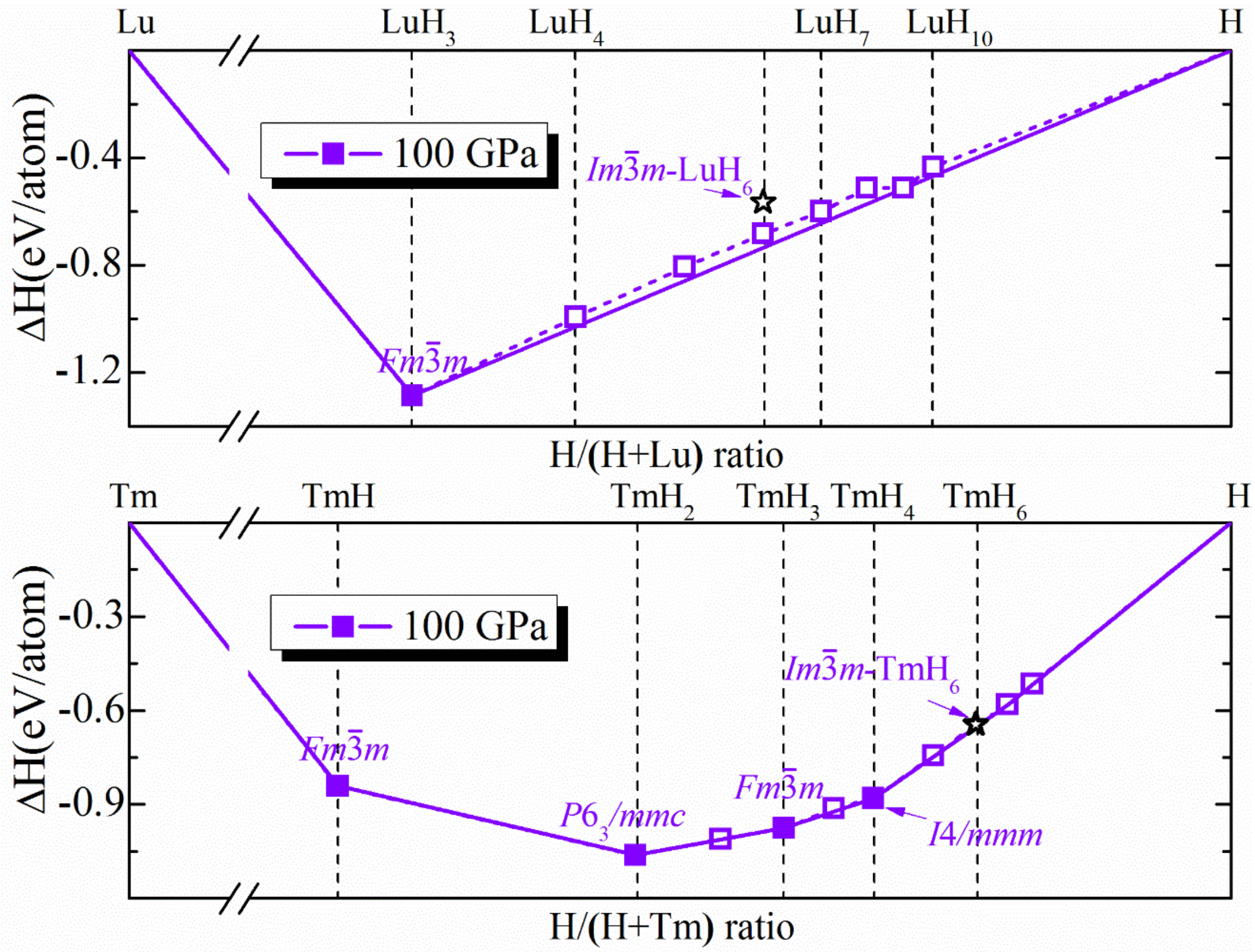

**Fig. S8** Phase diagrams of the Lu-H at 100 GPa and Tm-H at 50 GPa. The solid and open square violet symbols indicate that the structures are thermodynamic stable and unstable, respectively. The open black stars indicate the cubic $LuH_6$ (150 meV/atom above the convex hull) and $TmH_6$ (8 meV above the convex hull).

**Table S3.** Calculated structural parameters of Lu-H and Tm-H compounds.

| | Space group | Lattice parameters (Å) | Atoms | Atomic coordinates (fractional) | | |
|---|---|---|---|---|---|---|
| | | | | X | Y | Z |
| $LuH_3$ (100 GPa) | $Fm\bar{3}m$ | $a=b=c=4.35$<br>$\alpha=\beta=\gamma=90°$ | H(8c)<br>H(4b)<br>Lu(4a) | 0.25<br>0.50<br>0.00 | 0.25<br>0.50<br>0.00 | 0.25<br>0.50<br>0.00 |
| $LuH_6$ (100 GPa) | $Fm\bar{3}m$ | $a=b=c=3.69$<br>$\alpha=\beta=\gamma=90°$ | H(8c)<br>H(4b)<br>Lu(4a) | 0.25<br>0.50<br>0.00 | 0.25<br>0.50<br>0.00 | 0.25<br>0.50<br>0.00 |
| TmH (50 GPa) | $Fm\bar{3}m$ | $a=b=c=4.24$<br>$\alpha=\beta=\gamma=90°$ | H(4b)<br>Tm(4a) | 0.50<br>0.00 | 0.50<br>0.00 | 0.50<br>0.00 |
| $TmH_2$ (50 GPa) | $P6_3/mmc$ | $a=b=2.39$<br>$c=4.39$<br>$\alpha=\beta=\gamma=90°$ | H(2d)<br>H(2a)<br>Tm(2c) | 1/3<br>0.00<br>2/3 | 2/3<br>0.00<br>1/3 | 0.75<br>0.00<br>0.75 |
| $TmH_3$ (50 GPa) | $Fm\bar{3}m$ | $a=b=c=4.58$<br>$\alpha=\beta=\gamma=90°$ | H(8c)<br>H(4b)<br>Tm(4a) | 0.25<br>0.50<br>0.00 | 0.25<br>0.50<br>0.00 | 0.25<br>0.50<br>0.00 |
| $TmH_4$ (50 GPa) | $I4/mmm$ | $a=b=3.08$<br>$c = 5.40$<br>$\alpha=\beta=\gamma=90°$ | H(4d)<br>H(4e)<br>Tm(2a) | 0.50<br>0.50<br>0.00 | 0.00<br>0.50<br>0.00 | 0.25<br>0.094<br>0.00 |
| $TmH_6$ (50 GPa) | $Fm\bar{3}m$ | $a=b=c=3.85$<br>$\alpha=\beta=\gamma=90°$ | H(8c)<br>H(4b)<br>Tm(4a) | 0.25<br>0.50<br>0.00 | 0.25<br>0.50<br>0.00 | 0.25<br>0.50<br>0.00 |

## Electronic properties

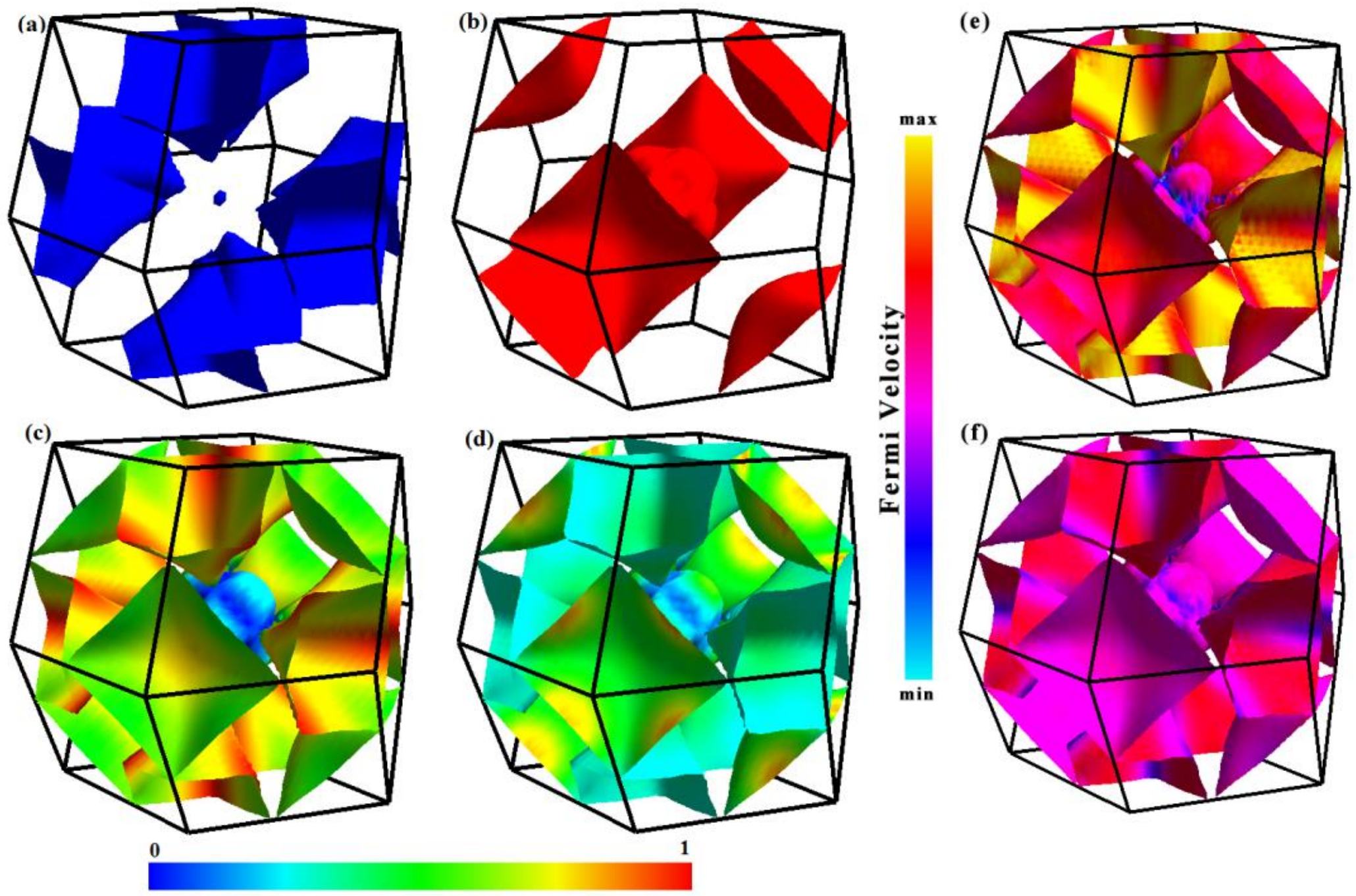

**Fig. S9.** The Fermi surface of cubic $YbH_6$ at 70 GPa. (a)-(b) The three-dimensional view of Fermi surface of the first and second band crossing the Fermi level. (c)-(d) Fermi surface with projection of 4*f* electrons of Yb and *s* electrons of H, respectively. (e)-(f) Fermi surface with projection of Fermi velocity with and without DFT+U (5 eV), respectively. The images were rendered using FermiSurfer software [19].

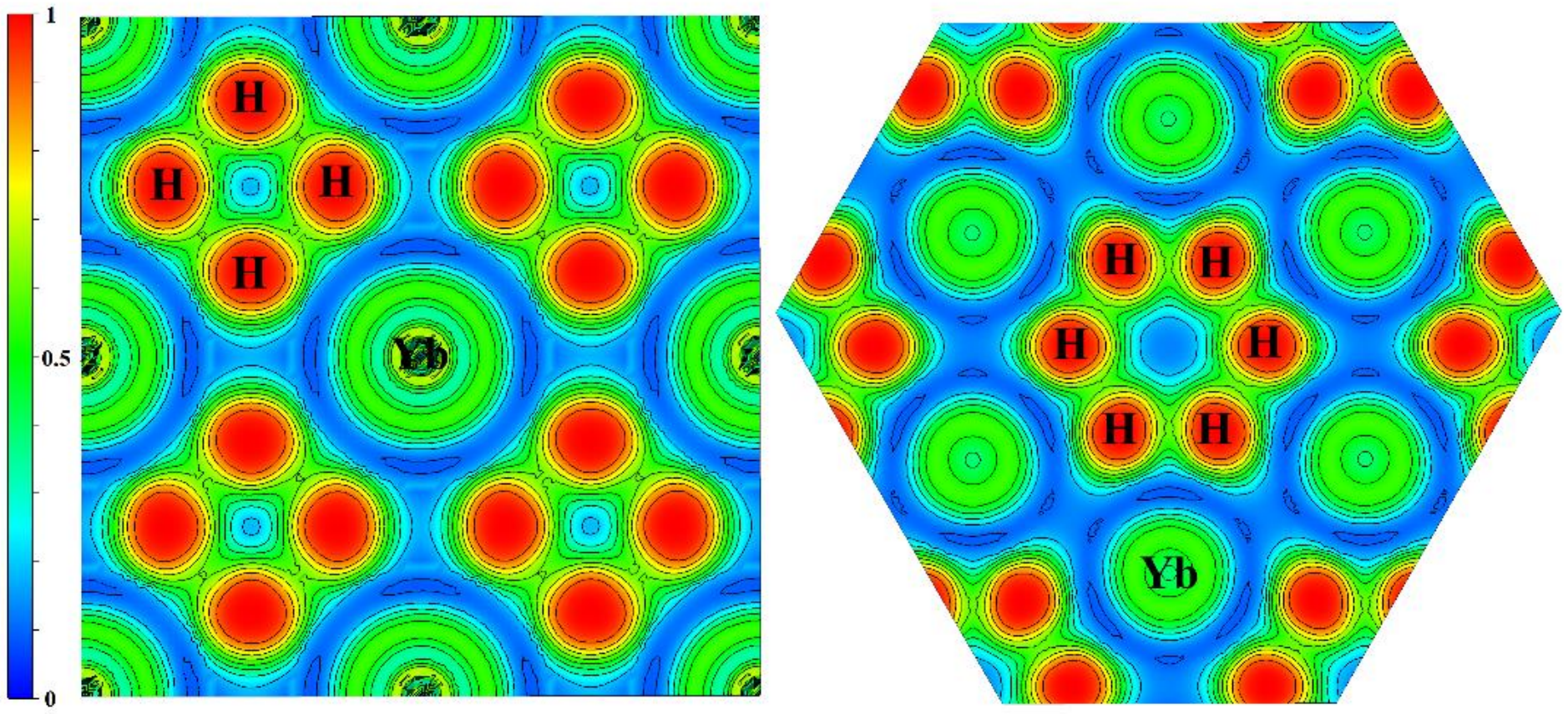

**Fig. S10.** The electron localization functions (ELF) of cubic $YbH_6$ at 70 GPa for [0 0 1] plane (left panel)

and [1 1 -1] plane (right panel). The isopleth curves plot with interval 0.1

## Equations for calculating $T_c$ and related parameters

### (1) Self-consistent iteration solution of the Eliashberg equation

The Migdal-Eliashberg equation has a form[20-22]:

$$\Delta(i\omega_n)Z(i\omega_n) = \frac{\pi T}{N_\mathrm{F}} \sum_{n'} \frac{\Delta(i\omega_n')}{\sqrt{\omega_n'^2 + \Delta^2(i\omega_n')}} \times [\lambda(\omega_n - \omega_{n'}) - N_\mathrm{F}\mu^*]\delta(\epsilon) \tag{S1}$$

$$Z(i\omega_n) = 1 + \frac{\pi T}{N_\mathrm{F}\omega_n} \sum_{n'} \frac{\omega_n'}{\sqrt{\omega_n'^2 + \Delta^2(i\omega_n')}} \lambda(\omega_n - \omega_{n'})\delta(\epsilon) \tag{S2}$$

Here functions $\Delta(i\omega_n)$ and $Z(i\omega_n)$ represent pairing order parameter and the renormalization function, $N_F$ is the density of electronic states at the Fermi level, and $\delta(\epsilon)$ is the Dirac delta function. $i\omega_n = i(2n+1)\pi T_c$ are the fermion Matsubara frequencies (we employ the themodynamic Green's functions formalism (see,e.g.,[23]); $\mu^*$ is the Coulomb pseudopotential, for which we use the widely accepted range of 0.1-0.13. $\lambda(\omega_n - \omega_{n'})$ contains the electron-phonon coupling matrix, phonon propagator. and the phonon density of states, and is given by:

$$\lambda(\omega_n - \omega_{n'}) = \int_0^\infty d\omega \frac{2\omega}{(\omega_n - \omega_n')^2 + \omega^2} \alpha^2 F(\omega) \tag{S3}$$

The equations for the order parameter and the renormalization function form a coupled nonlinear system and are solved self-consistently.We evaluated the renormalization function and the order parameter for each Matsubara frequency along the imaginary energy axis. After calculating $Z(i\omega_n)$ and $\Delta(i\omega_n)$, an analytic continuation is performed to the real axis using the Pade' functions.The calculation is performed for each T ($T_{min}$<$T$≤$T_{max}$) ($T_{min}$≈0 and $T_{max}$≥$T_c$). The critical temperature $T_c$ is obtained as an asymptotic value as $\Delta(i\omega_n)$ tends to zero.

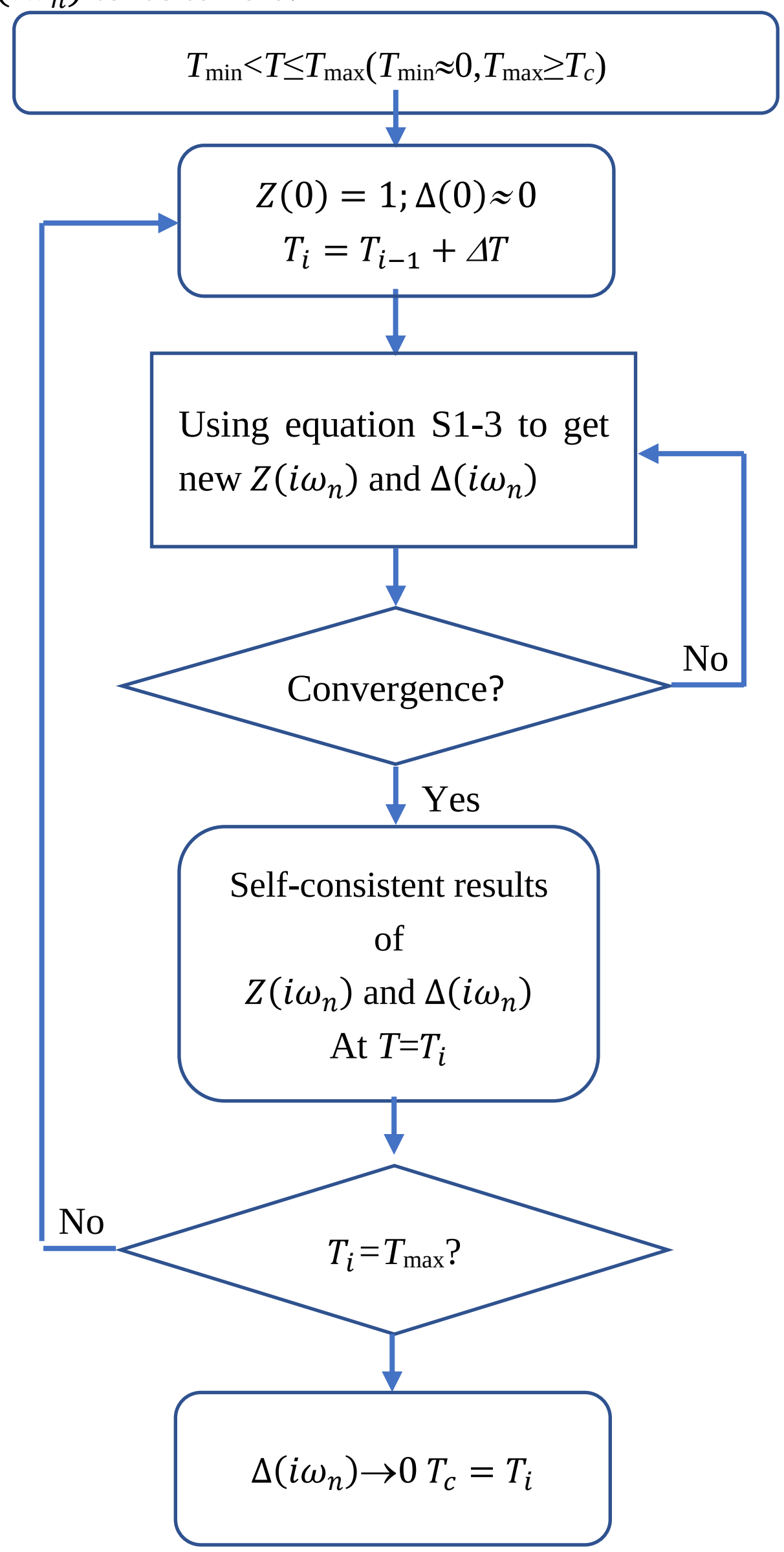

**(2) The Allen−Dynes-modified McMillan equation**

Allen-Dynes McMillan equation (A-D-M), which is the approximate analytic solution of the Eliashberg equations, at λ < 1.5 has a form: [11]:

$$T_c = \frac{\omega_{log}}{1.2} \exp\left[-\frac{1.04(1+\lambda)}{\lambda-\mu^*(1+0.62\lambda)}\right], \quad \text{(S4)}$$

If λ > 1.5, one should use the more general equation containing the corrections $f_1$ and $f_2$ :

$$T_c = \frac{f_1 f_2 \omega_{log}}{1.2} \exp\left[-\frac{1.04(1+\lambda)}{\lambda-\mu^*(1+0.62\lambda)}\right], \quad \text{(S5)}$$

where $f_1$ and $f_2$ are given by [11]:

$$f_1 = \sqrt[3]{\left[1+\left(\frac{\lambda}{2.46(1+3.8\mu^*)}\right)^{\frac{3}{2}}\right]}, f_2 = 1+\frac{(\frac{\bar{\omega}_2}{\omega_{log}}-1)\lambda^2}{\lambda^2+[1.82(1+6.3\mu^*)\frac{\bar{\omega}_2}{\omega_{log}}]^2}, \quad \text{(S6)}$$

Here $\bar{\omega}_2$ is themean square frequency,

$$\bar{\omega}_2 = \sqrt{\frac{2}{\lambda}\int \alpha^2 F(\omega)\, \omega d\omega}\,, \quad \text{(S7)}$$

$\omega_{log}$ is the logarithmic average frequency. The $\omega_{log}$ and EPC constant λ are defined by the relations:

$$\omega_{log} = \exp\left[\frac{2}{\lambda}\int \frac{d\omega}{\omega}\alpha^2 F(\omega)\ln\omega\right], \quad \text{(S8)}$$

$$\lambda = 2\int \frac{\alpha^2 \mathrm{F}(\omega)}{\omega} d\omega, \quad \text{(S9)}$$

**(3) Gor'kov-Kresin equation**

Gor'kov and Kresin (G-K) introduced the coupling constants $\lambda_{opt}$ and $\lambda_{ac}$ describing the interaction of electrons with optical and acoustic phonons[24,25]. The generalized Eliashberg equation has the form (at $T=T_c$):

$$\Delta(\omega_n)Z = \pi T \sum_{\omega_{n'}} \left[\lambda_{\mathrm{opt}} \frac{\tilde{\Omega}^2_{\mathrm{opt}}}{\tilde{\Omega}^2_{\mathrm{opt}}+(\omega_n-\omega_{n'})^2} + \lambda_{\mathrm{ac}} \frac{\tilde{\Omega}^2_{\mathrm{ac}}}{\tilde{\Omega}^2_{\mathrm{ac}}+(\omega_n-\omega_{n'})^2}\right] \frac{\Delta(\omega_{n'})}{|\omega_{n'}|}_{|T=T_c}, \quad \text{(S10)}$$

$$\lambda_{\mathrm{ac}} = 2\int_0^{\omega_1} \frac{\alpha^2 F(\omega)}{\omega} d\omega, \lambda_{\mathrm{opt}} = 2\int_{\omega_1}^{\omega_m} \frac{\alpha^2 F(\omega)}{\omega} d\omega, \lambda_{\mathrm{ac}} + \lambda_{\mathrm{opt}} = \lambda, \quad \text{(S11)}$$

where $\omega_1$ is the maximum frequency for the acoustic modes, $\omega_m$ is the maximum frequency's value. The mean square average frequency values are defined as follows:

$$\tilde{\omega}_{\mathrm{ac}} = \langle \omega_{\mathrm{ac}}{}^2 \rangle^{\frac{1}{2}}, \langle \omega_{\mathrm{ac}}{}^2 \rangle = \frac{2}{\lambda_{\mathrm{ac}}}\int_0^{\omega_1} d\omega \cdot \omega^2 \frac{\alpha^2 F(\omega)}{\omega} = \frac{2}{\lambda_{\mathrm{ac}}}\int_0^{\omega_1} \alpha^2 F(\omega)\omega d\omega, \quad \text{(S12)}$$

$$\tilde{\omega}_{\mathrm{opt}} = \langle \omega_{\mathrm{opt}}{}^2 \rangle^{\frac{1}{2}}, \langle \omega_{\mathrm{opt}}{}^2 \rangle = \frac{2}{\lambda_{\mathrm{opt}}}\int_{\omega_1}^{\omega_{\mathrm{m}}} d\omega \cdot \omega^2 \frac{\alpha^2 F(\omega)}{\omega} = \frac{2}{\lambda_{\mathrm{opt}}}\int_{\omega_1}^{\omega_{\mathrm{m}}} \alpha^2 F(\omega)\omega d\omega, \quad \text{(S13)}$$

For our predicted hydrides the $\lambda_{\mathrm{ac}} << \lambda_{\mathrm{opt}}$, we assume that:

$$T_c = T_c^{opt} + \Delta T_c^{ac}, \text{ and } T_c^{\mathrm{opt}} >> \Delta T_c^{\mathrm{ac}} \quad \text{(S14)}$$

As a result, the expression for Tc can be written in the form:

$$T_c = \left[1 + 2\frac{\lambda_{\mathrm{ac}}}{\lambda_{\mathrm{opt}} - \mu^*} \cdot \frac{1}{1+\rho^{-2}}\right] T_c^0, \rho = \frac{\tilde{\omega}_{\mathrm{ac}}}{\pi T_c^0}, T_c^0 \equiv T_c^{opt}. \quad \text{(S15)}$$

Here the $T_c^0$ is defined as the transition temperatures caused by the interaction of electrons with optical phonons only; for $\lambda_{\mathrm{opt}} \leq 1.5$:

$$T_c^0 = \frac{\tilde{\omega}_{\mathrm{opt}}}{1.2} \exp\left[-\frac{1.04(1+\lambda_{\mathrm{opt}})}{\lambda_{\mathrm{opt}} - \mu^*(1+0.62\lambda_{\mathrm{opt}})}\right]. \quad \text{(S16)}$$

For $\lambda_{\mathrm{opt}} > 1.5$:

$$T_c^0 = \frac{0.25\tilde{\omega}_{\mathrm{opt}}}{[e^{\frac{2}{\lambda_{eff}}} - 1]^{1/2}}. \quad \text{(S17)}$$

Here the $\lambda_{\mathrm{eff}}$ is defined as follows:

$$\lambda_{\mathrm{eff}} = (\lambda_{\mathrm{opt}} - \mu^*)\left[1 + 2\mu^* + \lambda_{\mathrm{opt}}\mu^* t(\lambda_{\mathrm{opt}})\right]^{-1}, \quad \text{(S18)}$$

$$t(x) = 1.5\exp(-0.28)\,x. \quad \text{(S19)}$$

We used the following expression:

$$\alpha = \frac{1}{2}[1 - 4\frac{\lambda_{ac}}{\lambda_{opt}}\frac{\rho^2}{(\rho^2+1)^2} \quad \text{(S20)}$$

to calculate the isotope coefficient α.

## Superconductive properties

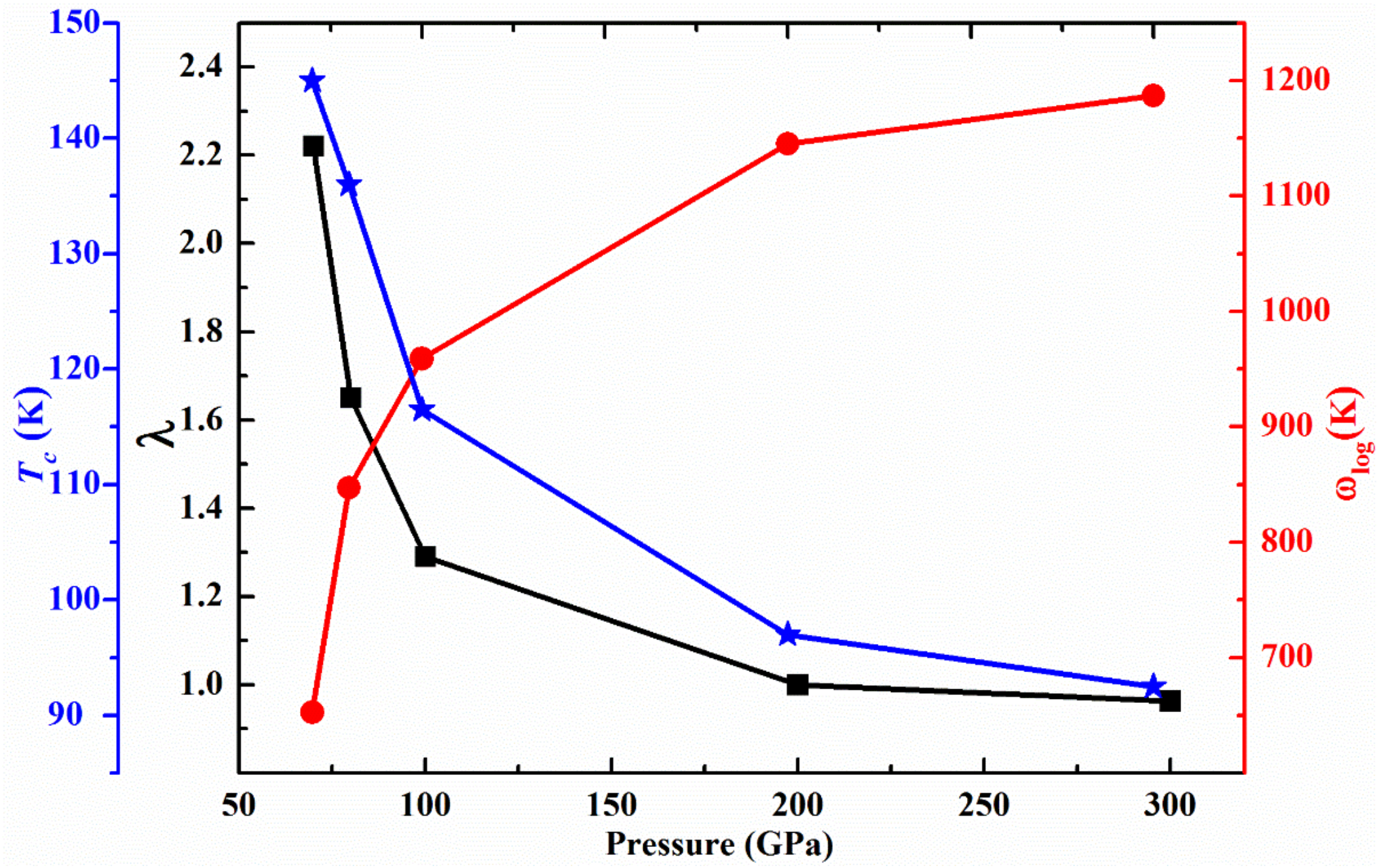

**Fig. S11.** The EPC parameter λ, logarithmic average frequency $\omega_{log}$, and $T_c$ of $YbH_6$ versus pressures

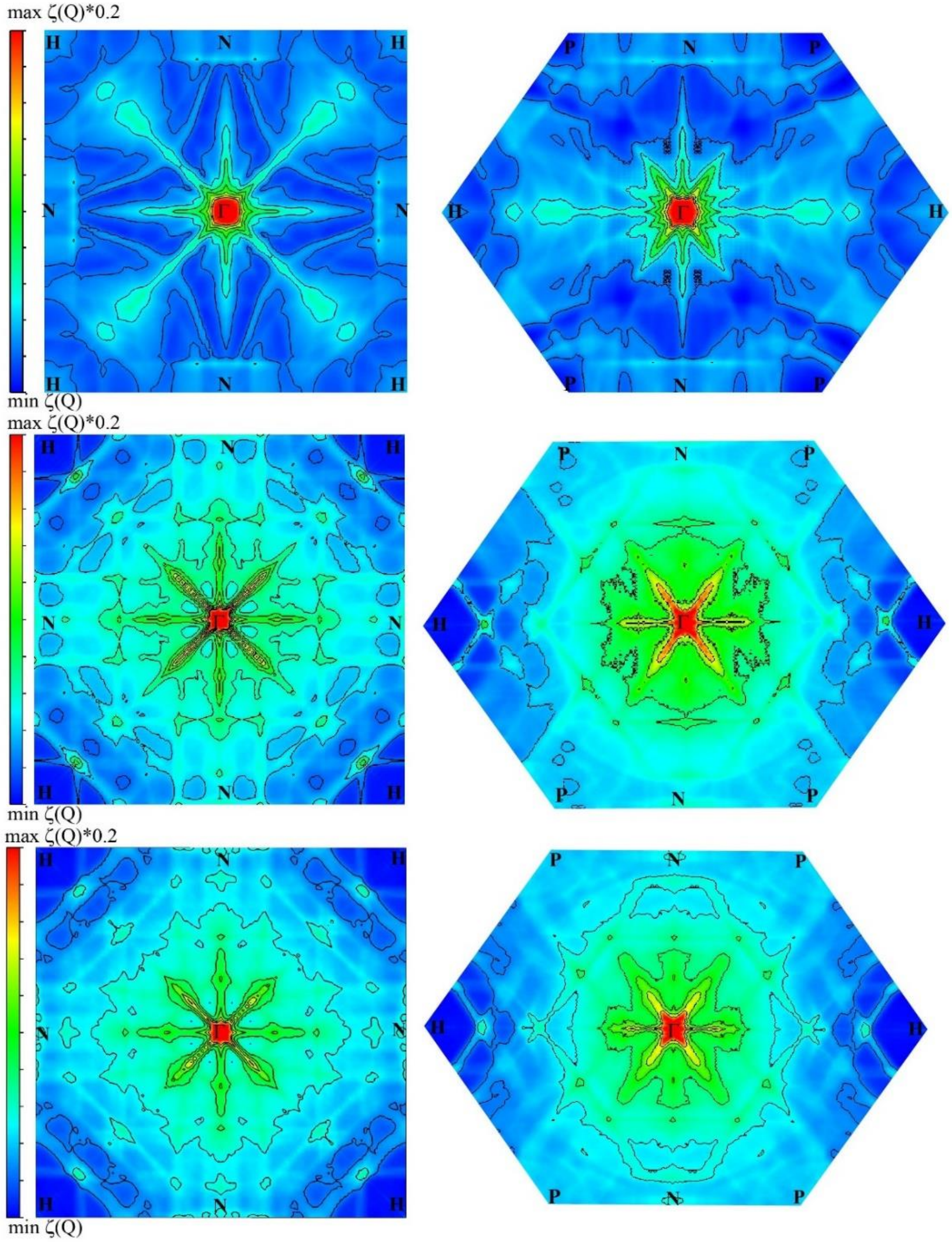

**Fig. S12.** The 2D Fermi nesting function of cubic $YbH_6$ at 70 GPa (up panel), $LuH_6$ at 100 GPa (middle panel) and $YH_6$ at 120 GPa (down panel) using ELK code with 100*100*100 k-points sampling. Strong nesting along F-H and H-N directions for cubic $YbH_6$. $LuH_6$ and $YH_6$ has quite similar Fermi nesting pattern.

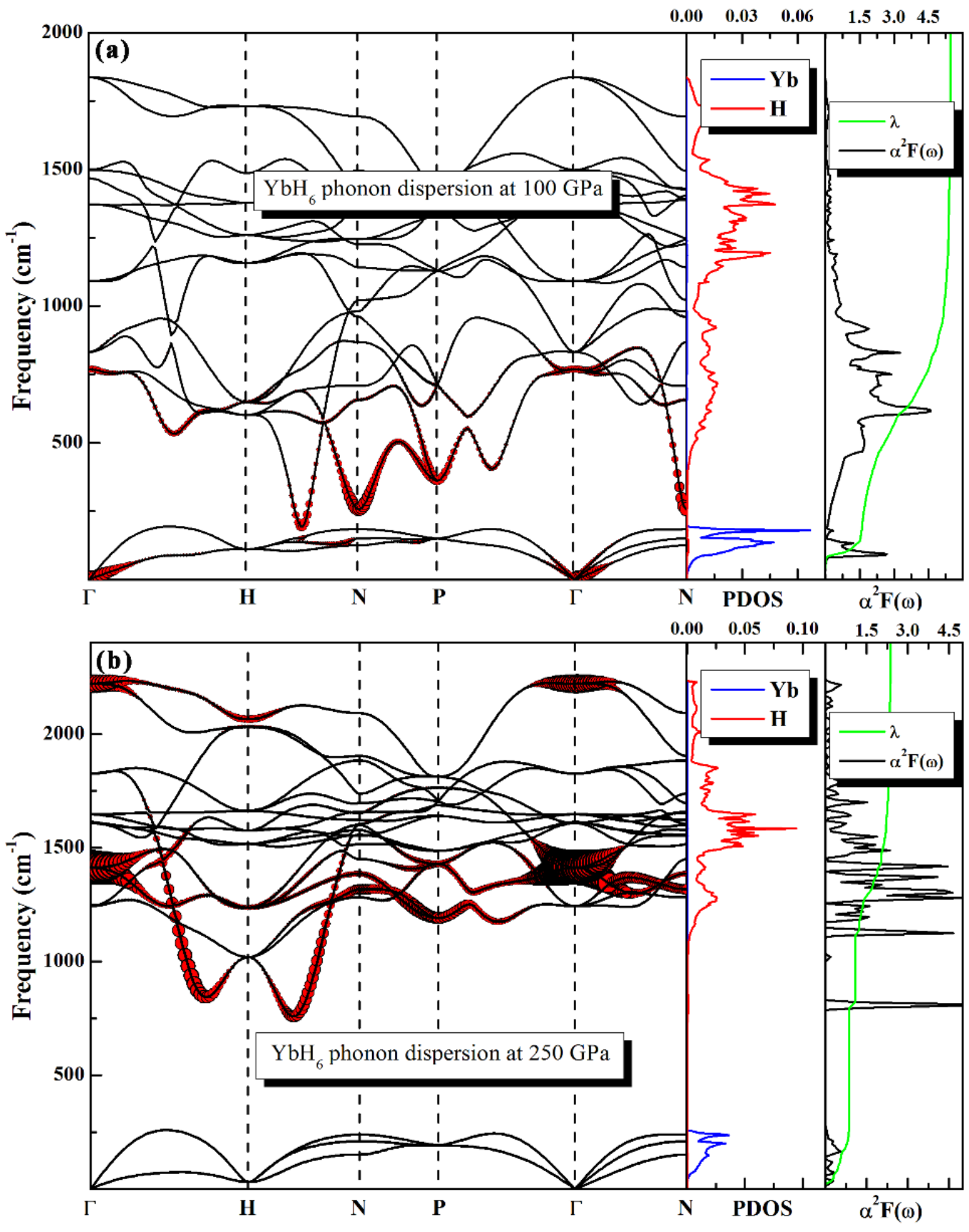

**Fig. S13.** Phonon dispersion, phonon density of state, spectral function ($\alpha^2F(\omega)$) and integral EPC $\lambda$ of *Fm*$\overline{3}$*m*-$YbH_6$ at 100 GPa (a) and at 250 GPa (b), with frozen-*f* electrons. Red circles with radius proportional to the EPC strength.

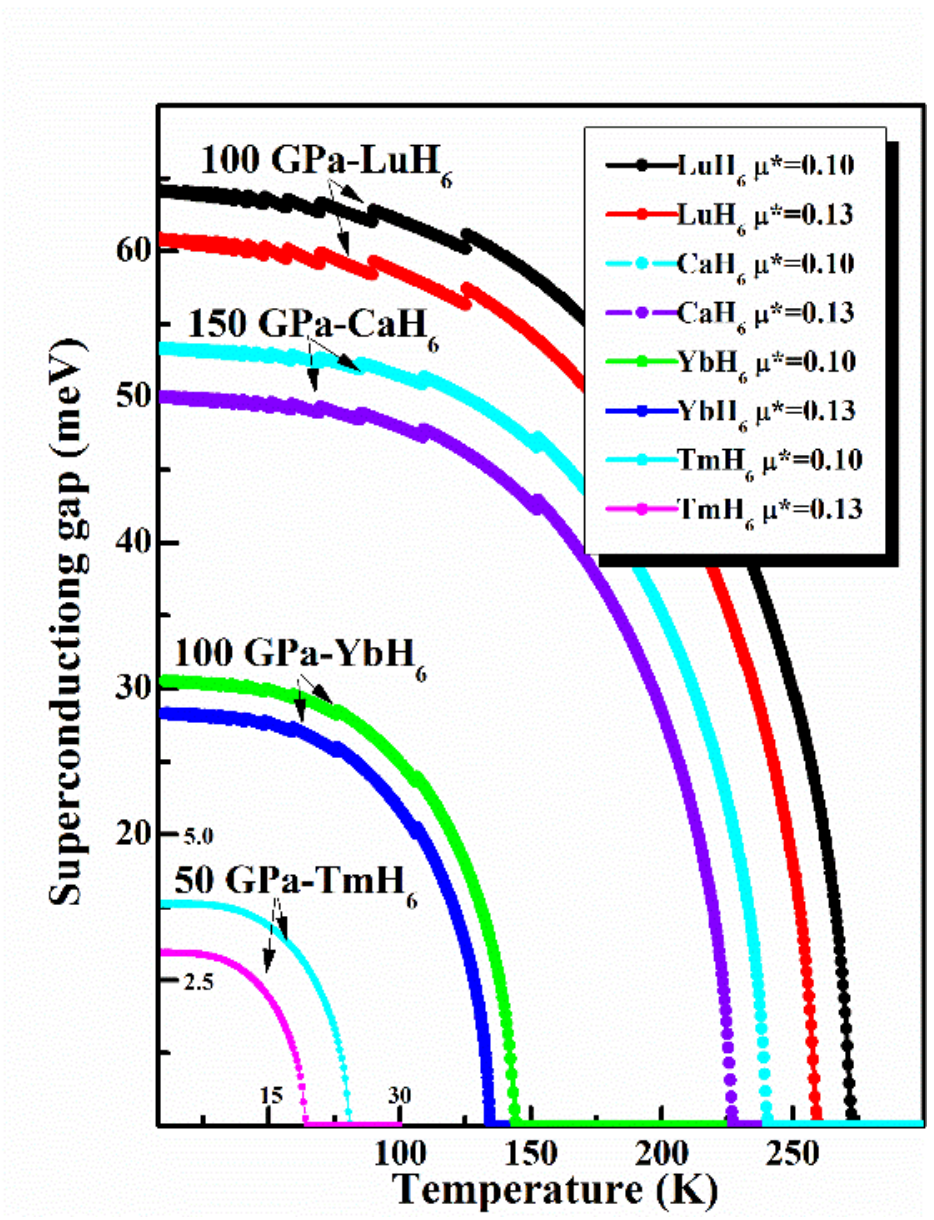

**Fig. S14.** The superconducting gap for $XH_6$ (X=Ca, Tm, Yb, Lu) at different pressure.

**Table S4.** The calculated EPC parameter $\lambda$, logarithmic average phonon frequency $\omega_{log}$ (K), electronic density of states at Fermi level $N(\varepsilon_f)$ (states/spin/Ry/f.u.) isotope coefficients $\alpha$ and superconducting transition temperatures $T_c$ (K) with $\mu^*$=0.1-0.13 at corresponding pressures $P$ (GPa).

| Structure | $P$ | $\lambda$ | $\omega_{log}$ | $N(\varepsilon_f)$ | $\alpha$ | $T_c$ (K)[a] | A-D $T_c$ (K)[b] | scE $T_c$ (K) | G-K $T_c$ (K) |
|---|---|---|---|---|---|---|---|---|---|
| $Im\bar{3}m$ $YbH_6$ | 70 | 2.22 | 652.4 | 8.4 | 0.48 | 90-94 | 114-125 | 125-145 | 121-131 |
| | 200 | 1.00 | 1146.5 | 6.8 | 0.38 | 69-80 | 72-85 | 82-96 | 78-91 |
| $Im\bar{3}m$ $LuH_6$ | 100 | 3.60 | 750.5 | 4.8 | 0.49 | 140-146 | 204-222 | 260-273 | 227-243 |
| | 200 | 1.77 | 1366.2 | 4.2 | 0.47 | 166-180 | 190-209 | 231-250 | 190-210 |
| $Im\bar{3}m$ $TmH_6$ | 50 | 0.72 | 611.9 | 29.6 | 0.47 | 17-22 | 18-23 | 19-24 | 19-25 |
| $Im\bar{3}m$ $EuH_6$ | 160 | 0.13 | 511.1 | 64.4 | \ | 0 | 0 | 0 | 0 |
| $Im\bar{3}m$ $CaH_6$ | 150 | 2.67 | 1027.1 | 2.4 | 0.38 | 165-175 | 205-222 | 227-241 | 220-236 |
| $Im\bar{3}m$ $ScH_6$ | 300 | 1.40 | 1233.4 | 5.0 | 0.30 | 119-132 | 132-148 | 155-175 | 168-187 |
| | 350 | 1.30 | 1330.8 | 4.8 | 0.30 | 118-131 | 129-146 | 148-167 | 164-184 |
| $Im\bar{3}m$ $YH_6$ | 120 | 3.06 | 829.0 | 4.7 | 0.46 | 143-150 | 195-213 | 244-257 | 218-234 |
| $Im\bar{3}m$ $H_3S$ | 200 | 1.88 | 1266.0 | 3.1 | 0.35 | 162-175 | 186-204 | 212-223 | 206-229 |

[a] $T_c$ was estimated using Allen-Dynes- McMillian equation with $f_1 f_2 = 1$ (Eq. S1).

[b] $T_c$ was estimated using Allen-Dynes- McMillian equation with $f_1 f_2 \neq 1$ (Eq. S2).

**Table S5.** The calculated $T_c$ of $XD_6$ (X=Yb, Lu, Y, Ca, and Sc) at different pressures.

| Parameter | $YbD_6$ | $LuD_6$ | $YD_6$ | $CaD_6$ | $ScD_6$ |
|---|---|---|---|---|---|
| | 70 GPa | 100 GPa | 120 GPa | 150 GPa | 300 GPa |
| $T_c^D$ (K) | 87-94 | 162-173 | 158-170 | 169-181 | 136-152 |